\documentclass[12pt]{article}

\newcommand{\be}{\begin{eqnarray}}
\newcommand{\ee}{\end{eqnarray}}

\newcommand{\np}{\newpage}

\def\so#1{{\rm SO}( #1)}
\def\su#1{{\rm SU}( #1)}
\def\sp#1{{\rm Sp}( #1)}
\def\u#1{{\rm U}( #1)}
\def\one{A}
\def\two{B}
\def\adag{  { a^\dagger_0}{} }
\def\badag{ {{\bar a}^\dagger_0}{} }
\def\vac{ | 0, p^+ \rangle }
\def\ret{\nonumber\\{}}
\def\la#1{\label{#1}}
\def\eqs#1{(\ref{#1})}
\def\half{\textstyle\frac{1}{2}}
\def\tf{\textstyle\frac{3}{4}}
\def\thrhalf{\textstyle\frac{3}{2}}

\font\cmss=cmss10 at 11pt \font\cmsss=cmss8 at 8pt

\def\inbar{\vrule height1.5ex width.4pt depth0pt}
\def\mininbar{\vrule height.75ex width.3pt depth0pt}
\def\cc{\relax\,\hbox{$\mininbar\kern-.2em{\hbox{\rm\tiny C}}$}}
\def\IZ{\relax\ifmmode\mathchoice
{\hbox{\cmss Z\kern-.4em Z}}{\hbox{\cmss Z\kern-.4em Z}}
{\lower.4pt\hbox{\cmsss Z\kern-.4em Z}}
{\lower1.2pt\hbox{\cmsss Z\kern-.4em Z}}\else{\cmss Z\kern-.4em Z}\fi}
\def\IC{\relax\,\hbox{$\inbar\kern-.3em{\rm C}$}}
\def\IR{\relax{\rm I\kern-.18em R}}

\newcommand{\1}{1\kern -3pt \mathrm{l}}
\newcommand{\PP}{\mathrm{I}\kern -2pt \mathrm{P}}

\newcommand{\bea}{\begin{eqnarray}}
\newcommand{\eea}{\end{eqnarray}}
\newcommand{\ba}{{\bar{a}}}
\newcommand{\bb}{{\bar{b}}}

\newcommand{\bw}{{\bar{w}}}

\newcommand{\al}{\alpha}
\newcommand{\ga}{\gamma}

\newcommand{\vphi}{\varphi}

\newcommand{\om}{\omega}
\newcommand{\Om}{\Omega}

\newcommand{\De}{\Delta}

\newcommand{\tha}{\theta}

\newcommand{\Ups}{\Upsilon}

\newcommand{\tOm}{\widetilde{\Omega}}
\def\tIR{\tilde{\IR}}
\newcommand{\Sp}{\mathrm{Sp}}
\newcommand{\SU}{\mathrm{SU}}
\newcommand{\SO}{\mathrm{SO}}
\newcommand{\U}{\mathrm{U}}

\newcommand{\R}{\mathrm{I}\kern -2.5pt \mathrm{R}}
\newcommand{\Z}{\mathsf{Z}\kern -5pt \mathsf{Z}}

\newcommand{\tr}{{\rm tr}}

\newcommand{\pa}{\partial}
\newcommand{\rar}{\rightarrow}
\newcommand{\non}{\nonumber}

\newcommand{\ds}{\displaystyle}
\newcommand{\diag}{\mathrm{diag}}

\newcommand{\cN}{\mathcal{N}}
\newcommand{\cO}{\mathcal{O}}
\newcommand{\cQ}{\mathcal{Q}}
\newcommand{\cA}{\mathcal{A}}
\newcommand{\tcQ}{\tilde{\mathcal{Q}}}
\newcommand{\tcA}{\tilde{\mathcal{A}}}

\def\Gori{G}
\def\Zori{\IZ_2^{\rm ori}}
\def\Zorb{\IZ_2^{\rm orb}}
\newcommand{\D}{{\rm d}}
\newcommand{\e}{{\rm e}}
\newcommand{\JR}{J_R}
\newcommand{\tA}{\tilde{A}}

\newcommand{\tQ}{\tilde{Q}}

\newdimen\tableauside\tableauside=1.0ex
\newdimen\tableaurule\tableaurule=0.4pt
\newdimen\tableaustep
\def\phantomhrule#1{\hbox{\vbox to0pt{\hrule height\tableaurule
width#1\vss}}}
\def\phantomvrule#1{\vbox{\hbox to0pt{\vrule width\tableaurule
height#1\hss}}}
\def\sqr{\vbox{%
  \phantomhrule\tableaustep

\hbox{\phantomvrule\tableaustep\kern\tableaustep\phantomvrule\tableaustep}%
  \hbox{\vbox{\phantomhrule\tableauside}\kern-\tableaurule}}}
\def\squares#1{\hbox{\count0=#1\noindent\loop\sqr
  \advance\count0 by-1 \ifnum\count0>0\repeat}}
\def\tableau#1{\vcenter{\offinterlineskip
  \tableaustep=\tableauside\advance\tableaustep by-\tableaurule
  \kern\normallineskip\hbox
    {\kern\normallineskip\vbox
      {\gettableau#1 0 }%
     \kern\normallineskip\kern\tableaurule}%
  \kern\normallineskip\kern\tableaurule}}
\def\gettableau#1 {\ifnum#1=0\let\next=\null\else
  \squares{#1}\let\next=\gettableau\fi\next}
\newcommand{\fund}{\tableau{1}}
\newcommand{\bfund}{\overline{\fund}}

\newcommand{\Yasymm}{\tableau{1 1}}
\newcommand{\basymm}{\overline{\tableau{1 1}}}
\begin{document}

\begin{flushright}
BRX-TH-504\\
BOW-PH-125\\
{\tt hep-th/0206094}
\end{flushright}
\vspace{.3in}
\setcounter{footnote}{0}
\stepcounter{table}

\begin{center}
{\Large{\bf \sf pp-wave limits and orientifolds}}

\vspace{.2in}

\begin{center}
Stephen G. Naculich$^a$,
Howard J. Schnitzer\footnote{Research
supported in part by the DOE under grant
DE--FG02--92ER40706.}$^{,b}$,
and Niclas Wyllard\footnote{
Research supported by the DOE under grant DE--FG02--92ER40706.\\
{\tt \phantom{aaa} naculich@bowdoin.edu;
schnitzer,wyllard@brandeis.edu}\\}$^{,b}$

\vspace*{0.3in}

$^{a}${\em Department of Physics\\
Bowdoin College, Brunswick, ME 04011}

\vspace{.2in}

$^{b}${\em Martin Fisher School of Physics\\
Brandeis University, Waltham, MA 02454}

\end{center}

\vskip 5mm

\end{center}

\begin{abstract}
We study the pp-wave limits of various elliptic models with orientifold planes 
and D7-branes, 
as well as the pp-wave limit of an orientifold of 
$\mathit{adS}_5{\times}T^{11}$.
Many of the limits contain both open and closed strings.
We also present pp-wave limits of theories which give rise to a 
compact null direction and contain  open strings.
Maps between the string theory states and gauge theory operators 
are proposed.
\end{abstract}

\np

\tableofcontents

\section{Introduction} \label{sint}

\setcounter{equation}{0}
\setcounter{footnote}{0}

Recently a particular limiting version of the adS/CFT
correspondence \cite{Maldacena:1998}
between string theory
on $\mathit{adS}_5{\times}S^5$ and the $\cN=4$ supersymmetric
gauge theory was discovered \cite{Berenstein:2002a}
(for earlier work
see refs.~\cite{Gueven:2000,Blau:2001,Metsaev:2001}).
On the string theory side the limiting metric is
the metric experienced by
an observer moving very fast along a particular null geodesic
of $\mathit{adS}_5{\times}S^5$,
viz., a parallel plane (pp) wave metric \cite{Penrose:1976}.
The limiting background also contains
a non-zero RR five-form.
We will refer to this limit
interchangeably as a Penrose or pp-wave limit.
Type IIB string theory in the pp-wave background is exactly
solvable \cite{Metsaev:2001,Metsaev:2002}.
Moreover, the  spectrum of string theory states in this background
corresponds to the subset of the 
single trace operators in the gauge theory
with large conformal dimension $\Delta$ and R-charge $J$, but with
$H \equiv \Delta-J$ finite.
The operators in the superconformal $\cN=4$ $\su{N}$
gauge theory that correspond to the string theory states
in the Penrose limit of $\mathit{adS}_5{\times} S^5$ were
identified in ref.~\cite{Berenstein:2002a}.

In this paper we study examples of the limiting procedure introduced
in ref.~\cite{Berenstein:2002a}
for a class of models containing orientifold planes 
(most of the models also contain D7-branes)
and therefore potentially having an open string sector.

Sections~\ref{sbac} and~\ref{spen} contain background
material relevant for our later discussions, with section \ref{spen}
exhibiting the two different classes of Penrose limits
encountered in this paper.
In section~\ref{sn=4} we consider the pp-wave
limit of the simplest orientifolded model,
$\mathit{adS}_5{\times}\IR\PP^5$ and its dual $\cN=4$ SO/Sp gauge
theory.
In section~\ref{sspa} we discuss some aspects of the simplest
orientifold of $\mathit{adS}_5{\times}S^5$ containing
D7-branes, whose dual is the $\cN=2$
$\sp{2N}$ gauge theory with $\Yasymm\oplus4\fund$ matter hypermultiplets.
The pp-wave limit containing an open string sector was
studied in ref.~\cite{Berenstein:2002b};
we focus on another Penrose limit of the same theory,
which does not contain open strings.
(Other aspects of open strings and D-branes in
the pp-wave background
have been discussed in
refs.~\cite{Open}.)

In sections~\ref{ssua} and~\ref{ssp1}
we discuss the pp-wave limits of theories
whose IIB descriptions
contain
(before taking the limit)
an O7-plane,
four D7-branes (plus mirror branes),
and a $\Z_2$ orbifold; specifically,
$\cN=2$ $\su{N}$ with $2\protect\Yasymm\oplus4\protect\fund$ matter
hypermultiplets,
and $\cN=2$ $\sp{2N}{\times}\sp{2N}$ with
$(\protect\fund,\protect\fund)\oplus
2(1,\protect\fund)\oplus 2(\protect\fund,1)$  matter hypermultiplets.
The adS/CFT correspondence for these elliptic models
was discussed in detail in refs.~\cite{Gukov:1998,
Ennes:2000}.
Both theories possess two different Penrose limits,
one of which allows for an open string sector.
Moreover, in these models there are several distinct
species of open strings,
due to the presence of the orbifold projection.

In sections~\ref{sgoose1} and \ref{sgoose2} we discuss two infinite
classes of elliptic models \cite{Park:1998}
for which the models considered in
sections~\ref{ssua} and~\ref{ssp1} are the first members.
The type IIB descriptions of these two classes of 
models contain an O7-plane,
four D7-branes (plus mirror branes),
and a $\Z_{2N_2}$ orbifold,
and the dual gauge theories have gauge groups $\SU^{N_2}$ and
$\Sp{\times} \SU^{N_2{-}1} {\times} \Sp$,
respectively.
We show that these theories possess a scaling limit, 
similar to the one recently considered in  refs.~\cite{Mukhi:2002, Alishahiha:2002b},
giving rise to a compact null direction.
The theories we consider contain both open and closed string sectors.

In section~\ref{ssp2} we discuss the pp-wave limit for the
correspondence between type IIB string theory on
$\mathit{adS}_5{\times}T^{11}/\Zori$ and the
$\cN=1$ $\Sp(2N){\times}\Sp(2N)$ theory with
$2 (\protect\fund,\protect\fund)\oplus
4(1,\protect\fund)\oplus 4(\protect\fund,1)$ matter chiral multiplets.
By studying this example we are able to test the ideas 
of ref.~\cite{Berenstein:2002b} in a more complicated setting.
Section \ref{ssum} summarizes the results of this paper.

\setcounter{equation}{0}
\section{Gauge theories on branes in orientifold backgrounds} \label{sbac}

The four-dimensional $\cN=4$, $\cN=2$, and $\cN=1$
supersymmetric gauge theories
considered in this paper arise as the worldvolume theories
on D3-branes in various orientifolded type IIB backgrounds.
The $\cN=4$ and $\cN=2$ theories arise from D3-branes in a flat background
modded out by an orientifold group $\Gori$.
We parameterize the flat six-dimensional space transverse to
the D3-brane worldvolume by
\bea
&&z_1= x_6+ix_7=r\cos{\theta_1}\cos{\theta_2}\,\e^{i\vphi_1},
\qquad\qquad
\theta_{1,2}\in [0,\half\pi],
\ret
&&z_2= x_8+ix_9=r\cos{\theta_1}\sin{\theta_2}\,\e^{i\vphi_2},
\qquad\qquad
\vphi_{1,2,3}\in [0,2\pi],
\ret
&&z_3= x_4+ix_5=r\sin{\theta_1}\,\e^{i\vphi_3}.
\la{param}
\ee
The coordinates $z_i$ are subject
to various identifications that depend on the form
of $\Gori$.

In the large-$N$ limit ($\sim$ large number of D3-branes),
the D3-branes significantly modify the background.
The near-horizon limit of the resulting background
is the space $\mathit{adS}_5{\times} S^5$,
whose metric is given by
\bea
\D s^2 &=&  R^2 \left( \D s_{\mathit{adS}_5}^2 + \D \tOm_5{}^2  \right) \,, \ret
\D s_{\mathit{adS}_5}^2 &=&
\D \rho^2 - \cosh^2 \rho \, \D t^2 +  \sinh^2 \rho \, \D \Omega_3^2
\label{adS5} \,,\\
\D \tOm_5{}^2  &=&
\D\theta_1{}^2 + \cos^2\theta_1 \,
\left(\D\theta_2{}^2
+ \cos^2\theta_2\,\D\vphi_1{}^2+\sin^2\theta_2\,\D\vphi_2{}^2\,\right)+
\sin^2\theta_1\, \D\vphi_3{}^2.
\la{s5}
\ee
Modding out by the orientifold group $\Gori$ results
in identifications among the coordinates, specified in each case below.

The isometry group of $S^5$ is $\so{6}$,
which is to be identified with the $\su{4}$ $R$-symmetry group
of the $\cN=4$ theories.
For the $\cN=2$ theories considered in this paper,
$\Gori$ reduces the isometry group to $[\so{4} \times \so{2}]/\Gori$,
where $\so{4} \simeq \su{2}_L\times\su{2}_R$
is the rotation group on $\IC^2\sim \{z_1, z_2\}$,
and $\so{2} \simeq \u{1}_R$ is the rotation group on $\IC\sim \{z_3\}$.
More precisely,
$\su{2}_L$ and $\su{2}_R$ act on the left and right of the matrix
\be
\pmatrix{   z_1 & -\bar{z}_2 \cr
            z_2 &  \bar{z}_1 } \,.
\ee
The diagonal generators $J^3_{L,R}$ of $\su{2}_{L,R}$
and the generator $J_R$ of $\u{1}_R$, act as
\be
J^3_R + J^3_L = -i\partial_{\vphi_1}\,, \qquad
J^3_R - J^3_L = -i\partial_{\vphi_2}\,, \qquad
          \JR = -i\partial_{\vphi_3}\,.
\la{Jident}
\ee
$\su{2}_R\times\u{1}_R/\Gori$ is the $R$-symmetry group
of the $\cN=2$ field theories,
whereas $\su{2}_L$ is a global symmetry group.
For some of the models we consider $\SU(2)_L$ is further reduced to $\U(1)_L$.
For these models $J^3_L$ is identified with the $\U(1)_L$ generator.
To facilitate comparison to the results in ref.~\cite{Ennes:2000},
we note that $J_R^3 = \half q_R$.

In this paper, we also consider the 
$\cN=1$ $\Sp(2N){\times}\Sp(2N)$ field
theory on a stack of D3-branes in the background of an orientifold
of the conifold. The orientifold group and the gauge theory 
on the D3-branes are discussed
in more detail in section~\ref{ssp2}.

\section{Penrose limits of orientifolds of $\mathit{adS}_5{\times}S^5$}
\label{spen}

All geodesics of $S^5$ are equivalent,
so there is only one Penrose limit of $\mathit{adS}_5{\times} S^5$,
when the geodesic lies partially in the $S^5$ directions.
(If the geodesic lies completely inside $\mathit{adS}_5$,
the Penrose limit gives flat space \cite{Blau:2002b}.)
For the orientifolded theories considered in this paper,
however, Penrose limits along different geodesics can give rise
to different string theories.
In this section, we identify two particular geodesics
and the associated Penrose limits.

\bigskip
\noindent{\bf Penrose limit \one}

\noindent Consider the geodesic
parameterized by $\vphi_3$
and contained in the subspace $\rho=0$, $\theta_1 = \half\pi$.
Define
\be
x^+ = \half (t + \vphi_3)\,, \qquad
x^- = \half R^2 (t - \vphi_3)\,, \qquad
r   = R \rho, \qquad
y   = R(\half \pi - \theta_1)\,.
\la{Penrose1}
\ee
In the limit $R \to \infty$ (keeping $x^\pm$, $r$,  and $y$ finite),
the metric goes to
\be
\D s^2 = -4 \D x^+ \D x^- - (r^2 + y^2) (\D x^+)^2
 + \D r^2 + r^2 \D \Om_3^2
 + \D y^2 + y^2 \D {\tOm}_3^2\,,
\label{ppone}
\ee
where
\be
\D \tOm_3^2=
\D\theta_2{}^2 + \cos^2\theta_2\,\D\vphi_1{}^2+\sin^2\theta_2\,
\D\vphi_2{}^2\,.
\ee
The space transverse to the null geodesic is $\IR^4 \times \tIR^4$.
Parameterizing $\tilde{\IR}^4 = \IC^2$ by
\be \label{w1w2}
w_1= y\cos \theta_2 \e^{i\vphi_1}, \qquad
w_2= y\sin \theta_2 \e^{i\vphi_2},
\ee
we can rewrite the metric as
\be
\D s^2 = -4 \D x^+ \D x^- - (r^2 + |w_1|^2 + |w_2|^2) (\D x^+)^2
 + \D r^2 + r^2 \D \Om_3^2 + |\D w_1|^2 + |\D w_2|^2 .
\la{Penone}
\ee
The $p^-$ lightcone momentum is given by 
$2p^- = H_\one =i\partial_{x^+} = i(\partial_t + \partial_{\vphi_3})$, 
which is identified in the dual gauge theory with \cite{Berenstein:2002a} 
$H_\one = \Delta- J_\one$, where $\Delta$ is the conformal dimension and 
$J_\one$ is a global symmetry generator. 
{}From \eqs{Jident} we see that
$J_\one = \JR$.
Hence, in Penrose limit  \one,
$H_\one = \Delta- \JR$.

\bigskip
\noindent{\bf Penrose limit \two}

\noindent Consider
the geodesic
parameterized by $\vphi_1$
lying along $\rho=0$ and $\theta_1 = \theta_2 = 0$.
Define
\be
x^+ = \half (t + \vphi_1)\,, \qquad
x^- = \half R^2 (t - \vphi_1)\,, \qquad
r   = R \rho\,, \qquad
y_2   = R \theta_2\,, \qquad
y_3   = R \theta_1\,.
\la{Penrose2}
\ee
In the limit $R \to \infty$ (keeping $x^\pm$, $r$, $y_2$, and $y_3$ finite),
the metric goes to
\be
\D s^2 = -4 \D x^+ \D x^- \!- (r^2 {+}\, y_2^2 {+}\, y_3^2) (\D x^+)^2
 \!+ \D r^2 \!+ r^2 \D \Om_3^2
 + \D y_2^2 + y_2^2 \D \vphi_2^2
 + \D y_3^2 + y_3^2 \D \vphi_3^2.
\la{pptwo}
\ee
The space transverse to the null geodesic is $\IR^4 \times \tilde{\IR}^4$.
Parameterizing $\tilde{\IR}^4 = \IC^2$ by
\be \label{w2w3}
w_2= y_2 \e^{i\vphi_2},\qquad
w_3= y_3 \e^{i\vphi_3},
\ee
the metric becomes
\be
\D s^2 = -4 \D x^+ \D x^- - (r^2 + |w_2|^2 + |w_3|^2) (\D x^+)^2
 + \D r^2 + r^2 \D \Om_3^2 + |\D w_2|^2 + |\D w_3|^2\,.
\la{Pentwo}
\ee
{}From $2p^- = H_\two = i(\pa_t+\pa_{\vphi_1})$ we see that 
$J_\two  = -i\partial_{\vphi_1}$, which using \eqs{Jident}, implies that
$J_\two = J_R^3+J_L^3$.
Hence, in Penrose limit  \two,
$H_\two = \Delta- J_R^3-J_L^3$.

The Penrose limits discussed here are limits of the full
supergravity/string theory background 
(as discussed in ref.~\cite{Gueven:2000}).
The non-trivial RR five-form field strength is also non-zero
in the limit (and is responsible for giving masses to the
lightcone fermions).
As observed in ref.~\cite{Berenstein:2002a},
string theory in the pp-wave background
is dual to a certain subset of gauge theory operators.

The metrics in the two Penrose limits
\eqs{Penone} and \eqs{Pentwo}
are identical.
However, the orientifold group $\Gori$ imposes different identifications
on the coordinates in the two cases, and thus can lead
to distinct string theories.
These distinct Penrose limits therefore correspond
to different subsets of operators in the gauge theory
on the D3 branes in the orientifolded background.

Finally, we note that the limit along the geodesic parameterized by $\vphi_2$
is equivalent to the limit involving $\vphi_1$ (Penrose limit $\two$).
On the gauge theory side, the only difference
is a trivial sign redefinition of $J^3_L$.

In section~\ref{ssp2} we discuss the Penrose limit of an orientifold of $\mathit{adS}_5{\times}T^{11}$.
More details about the orientifold group and the Penrose limit are given in that section.

\setcounter{equation}{0}
\section{$\cN=4$ $\sp{2N}$ and $\so{N}$ gauge theories} \label{sn=4}

Let us start by considering the
$\cN=4$ $\so{N}$ and $\sp{2N}$ theories.
In $\cN=1$ language, these theories contain a vector multiplet,
and three chiral multiplets ${\phi_{ia}}^b$.
The scalars in the chiral multiplets, organized into $SU(2)_R$
multiplets, and their quantum
numbers are shown in the table below.

\vbox{
\begin{center}
\begin{tabular}{|c|c|c|c|c|c|c|c|}
\hline
CFT field & $\sp{2N}$ or $\so{N}$ &
$\JR$ & $J_R^3$ & $J_L^3$ &
$\Delta$ & $H_\one$ & $H_\two$ \\
\hline
& & & & &  & & \\[-13pt]
\hline
& & & & &  & & \\[-13pt]
$\phi_3$     		& adjoint & 1  & 0       &        0 & 1 & 0 & 1 \\
$\phi_3^\dagger$     	& adjoint & $-1$ & 0       &        0 & 1 & 2 & 1 \\
$(\phi_1,\phi_2^\dagger)$     & adjoint & 0 & $(\half,-\half)$ &  $\half$ & 1 & (1,1) & (0,1) \\[1pt]
$(\phi_2,\phi_1^\dagger)$     & adjoint & 0 & $(\half,-\half)$ & $-\half$ & 1 & (1,1) & (1,2) \\[3pt]
\hline
\end{tabular}
\end{center}
\begin{center}
{\noindent\footnotesize{\bf Table \arabic{table}:\stepcounter{table}}
$\su{2}_R$ multiplets of scalar fields
in $\cN=4$ $\sp{2N}$ or $\so{N}$.}
\end{center}
}

The $\cN=4$ $\so{N}$ and $\sp{2N}$ gauge theories
arise as the worldvolume theories on a stack of D3-branes
parallel to an O3-plane in a flat type IIB background.
The generator $\Omega (-1)^{F_L}R_{456789}$
of the orientifold group $\Zori$
acts on the $\IC^3$ transverse
to the D3-branes as
\be
z^1 \to -z^1 , \qquad
z^2 \to -z^2 , \qquad
z^3 \to -z^3 ,
\la{o3action}
\ee
where $R_{456789}$ denotes reflection in the directions $4, \cdots, 9$,
$\Omega$ is the worldsheet parity operation, and
$(-1)^{F_L}$ changes the sign of the left-movers in the Ramond sector.

In the near-horizon limit,
the geometry becomes $\mathit{adS}_5 {\times} S^5/\Zori$
where the geometric part of $\Zori$ acts as
\be
\vphi_1 \to \vphi_1 + \pi \,,\qquad
\vphi_2 \to \vphi_2 + \pi \,,\qquad
\vphi_3 \to \vphi_3 + \pi \,,
\la{o3angle}
\ee
on the coordinates of $S^5$ (\ref{s5}), which yields
$\mathit{adS}_5 {\times} \IR \PP^5$. The adS/CFT correspondence
between string theory on this background and the above SO/Sp gauge
theories was developed in ref.~\cite{Witten:1998b}. 
The Penrose limit of these theories was briefly studied
in ref.~\cite{Floratos:2002}; below we give some further details.

\bigskip
\noindent{\bf Penrose limit \one}
\medskip

\noindent The action \eqs{o3angle} of $\Zori$ on the coordinates
in Penrose limit \one~is an inversion of $\tIR^4$:
$ w_1 \to -w_1$, $ w_2 \to -w_2$,
accompanied by a shift
$\vphi_3 \to \vphi_3 + \pi$
halfway around the geodesic (see section \ref{spen} for
details about the notation).
This $\Zori$ action has no fixed points on $S^5$,
and therefore no fixed points in the Penrose limit 
(as we will see in later sections, this fact is
related to the absence of an open string sector).
We now discuss the operators in the gauge theory that
correspond to the states in the closed string theory.

The only candidate for the operator corresponding to the vacuum state
of the closed string theory in the pp-wave background \eqs{Penone}
is\footnote{Throughout this paper we will not keep track of the
overall normalizations of the operators.}
\be
\tr \left[   \phi_3^{J_\one} \right] ,
\ee
which has $\Delta = J_\one$ and therefore $H_\one=0$.
The effect of modding out by  $\Zori$
is to halve the periodicity of the geodesic circle,
and thus to allow only even integer values of $J_\one$
for the vacuum string states. Thus from the string theory perspective one
expects the vacuum to correspond to
\be
\la{adjointvac}
\tr \left[   \phi_3^{2n} \right] .
\ee
{}From the gauge theory perspective, the exponent
must be even because the adjoint fields
satisfy $\phi^T_i =J \phi_i J$
($\phi_i^T = - \phi_i$) in the $\Sp$ theory ($\SO$ theory),
so that $\tr(\phi_3^{2n+1})$ vanishes identically.

Consider string states in which a zero-momentum
oscillator $\adag^\mu$ in one of the directions of $\IR^4$
acts on the light-cone vacuum.
These states correspond to the insertion of $(D_\mu \phi_3) \phi_3$
into \eqs{adjointvac}, yielding four operators
\be
\la{four}
\tr \left[  (D_\mu \phi_3) \phi_3^{2n+1} \right] ,
\ee
which have $\Delta=2n+3$ and $J_\one=2n+2$, and therefore $H_\one=1$.

Next, consider string states in which a zero-momentum
oscillator $\adag^i$, $\badag^i$ in one of the directions $w_i$, $\bw_i$
($i=1,2$)
of $\tIR^4$ acts on the light-cone vacuum.
The inversion $ (w_1, w_2)  \to (-w_1,-w_2)$ causes
$\adag^i \to -\adag^i$, $\badag^i \to -\badag^i$,
so for the string state to be invariant under $\Zori$,
the value of $J_\one$ must be odd.
These string states correspond to insertions of
$\phi_1 \phi_3$, $\phi_1^\dagger \phi_3$,
$\phi_2\phi_3$, and $\phi_2^\dagger\phi_3$ into (\ref{adjointvac}), 
yielding four additional $H_\one=1$ operators
\be
\la{fourmore}
\tr \left[  \phi_{1}\phi_3^{2n+1} \right],\qquad
\tr \left[  \phi^\dagger_{1}\phi_3^{2n+1} \right],\qquad
\tr \left[  \phi_{2}\phi_3^{2n+1} \right],\qquad
\tr \left[  \phi^\dagger_{2}\phi_3^{2n+1} \right].\qquad
\ee
Again, from the gauge theory perspective,
the exponents follow from the Sp (SO) constraints on the $\phi_i$'s.

The eight operators \eqs{four} and \eqs{fourmore} therefore
represent the eight bosonic zero-momentum states in the string theory.
The correspondence for the fermionic oscillators as well as for 
the non-zero-momentum oscillators 
can be constructed in complete analogy with the original
construction \cite{Berenstein:2002a};
we will not give the details here.

For the case of $\mathit{adS}_5{\times} \IR \PP^5$,
the distinction between
the $\Sp$ and $\SO$ theories was explained in ref.~\cite{Witten:1998b}.
An interesting question is to understand how they are distinguished in
the pp-wave limit.

\bigskip
\noindent{\bf Penrose limit \two}
\medskip

\noindent The action \eqs{o3angle} of $\Zori$ on the coordinates
in Penrose limit \two~involves an inversion 
of the transverse $\IC^2$:
$ w_2 \to -w_2$,
$ w_3 \to -w_3$,
accompanied by a shift
$\vphi_1 \to \vphi_1 + \pi$
halfway around the geodesic. 
This action is identical to the one in 
Penrose limit \one{} after a trivial
relabelling of the coordinates.
The correspondence of string states to the gauge
theory operators is the same as in Penrose limit \one{} except
with $\phi_1$ and $\phi_3$ interchanged.

\setcounter{equation}{0}
\section{$\cN=2$ $\sp{2N}$ with $\protect\Yasymm\oplus4\protect\fund$}
\la{sspa}

The $\cN=2$ $\sp{2N}$ theory with one antisymmetric
and four fundamental hypermultiplets
contains an $\cN=2$ vector multiplet in the adjoint representation,
which in $\cN=1$ language
consists of a vector multiplet
and a chiral multiplet, ${\phi_a}^b$,
in the adjoint representation.
The theory also contains an antisymmetric $\cN=2$ hypermultiplet,
which in $\cN=1$ language
consists of a chiral multiplet in
the $\Yasymm$ representation of $\Sp(2N)$, $A_{ab} = -A_{ba}$,
and a chiral multiplet in
the $\overline{\Yasymm}$ representation $\tA^{ab} = -\tA^{ba}$.
The indices on the latter can be lowered using the
symplectic unit $J_{ab}$,
leaving us with two chiral multiplets,
$A_{1ab}=A_{ab}$ and
$A_{2ab} = J_{ac} J_{bd} \tA^{ab}$,
in the $\Yasymm$ representation,
transforming as a doublet of $\su{2}_L$.
Finally,
the theory contains four fundamental $\cN=2$ hypermultiplets,
which in $\cN=1$ language
consists of eight chiral multiplets $Q^I_a$ in the $\fund$ representation
(after lowering the indices of the four chiral multiplets in the $\bfund$
representation).
The scalars in the chiral multiplets and their quantum
numbers are shown in the table below.

\vbox{
\begin{center}
\begin{tabular}{|c|c|c|c|c|c|c|c|c|}
\hline
CFT field & $\sp{2N}$ &
$\JR$ & $J_R^3$ & $J_L^3$ &
$\Delta$ & $H_\one$ & $H_\two$ \\
\hline
& & & & &  & & \\[-13pt]
\hline
& & & & &  & & \\[-13pt]
$\phi$     & adjoint   & 1 & 0       &        0 & 1 & 0 & 1 \\
$\phi^\dagger$     & adjoint   & $-1$ & 0       &        0 & 1 & 2 & 1 \\
$(A_1, (J A_2 J)^\dag)$& $\Yasymm$ & 0 & $(\half,-\half)$ &  $\half$ & 1 & (1,1) &(0,1) \\
$(A_2,(J A_1 J)^\dag)$ & $\Yasymm$ & 0 & $(\half,-\half)$ & $-\half$ & 1 & (1,1) &(1,2) \\
$(Q^I,Q^{I\dagger})$  & $\fund$ & 0 & $(\half,-\half)$ &        0 & 1 & 1 & $(\half,\frac{3}{2})$ \\[3pt]
\hline
\end{tabular}
\end{center}
\begin{center}
{\noindent\footnotesize{\bf Table \arabic{table}:\stepcounter{table}}
$\su{2}_R$ multiplets of scalar fields
in $\cN=2$  $\sp{2N}$\,\,with\,\,$\Yasymm\,\oplus\,4\,\fund$\,.}
\end{center}
}

The above $\cN=2$ $\sp{2N}$ gauge theory
arises~\cite{Sen:1996} as the theory
on a stack of D3-branes parallel to an O7-plane and four 
physical D7-branes in a flat background.
The orientifold group
$\Zori = \{\1 , \Omega R_{45}(-1)^{F_L}\}$,
where $R_{45}$ reflects the 45 directions,
acts on the space transverse to the D3-branes as
$z^3 \to -z^3$.
This action fixes the hyperplane $z_3=0$,
which corresponds to the position of the O7-plane
and the D7-branes.

In the near-horizon limit the background 
becomes $\mathit{adS}_5 {\times} S^5/\Zori$,
where $\Zori$ acts as $\vphi_3 \to \vphi_3 + \pi$
on the coordinates of $S^5$.
The fixed point set of this action is an $S^3$ located at $\theta_1=0$.
The adS/CFT correspondence for this model was discussed
in ref.~\cite{Fayyazuddin:1998}.

\bigskip
\noindent{\bf Penrose limit \one}
\medskip

\noindent In Penrose limit \one,
the generator of $\Zori$ simply produces a translation
along the geodesic $\vphi_3 \to \vphi_3 + \pi$
(together with the action of $\Omega (-1)^F$)
with no action on the transverse coordinates $w_1$ and $w_2$ (\ref{w1w2}).
This limit therefore yields the maximally supersymmetric
pp-wave background.
However, $\Zori$ projects out half of the discrete values
of $J_\one$ that would be allowed in the unprojected case.

\medskip
\noindent{\it Closed string sector}
\medskip

\noindent The $H_\one=0$ operators
\be
\tr \left[   {\phi^{2n}} \right],
\ee
correspond to the closed string vacuum states,
where the effect of modding out by $\vphi_3 \to \vphi_3 + \pi$
is to allow only even integer values of $J_\one$.
{}From the gauge theory perspective, the exponent must
be even because the adjoint field satisfies $\phi^T = J \phi J$,
implying that $ \tr \left[   {\phi^{2n+1}} \right]$ vanishes identically.

String states in which a zero-momentum oscillator
$\adag^\mu$ in one of the directions of $\IR^4$
acts on the light-cone vacuum
correspond to the $H_\one=1$ gauge theory operators
\be
\tr \left[  (D_\mu \phi) \phi^{2n+1} \right].
\ee

String states in which a zero-momentum oscillator
$\adag^i$, $\badag^i$ in one of the directions of $\tIR^4$
acts on the light-cone vacuum
correspond to the four $H_\one=1$ operators
\be
\tr \left[  (A_i J)  \phi^{2n} \right], \qquad
\tr \left[  (J A^\dagger_i)  \phi^{2n} \right], \qquad (i=1,2)\,,
\ee
where we write $A_i J$ for $A_{iac} J^{cb}$
 (and similarly for $J A_i^\dagger $).
Using $ (A J)^T = - J (A J) J$,
it follows that the operator would vanish if the exponent
of $\phi$ were odd.

Non-zero-momentum states can be constructed in complete analogy with the 
original construction \cite{Berenstein:2002a}.

The fixed point set (O7-plane) of $\Zori$ goes away in Penrose limit \one,
and so do the D7-branes,
so there is no open string sector.
The absence of an open string sector corresponds
to the fact that the operator
\be
Q^{I}  J \phi^n Q^{K}
\ee
vanishes by using the F-term equations of the gauge theory.
This implies that it is not a chiral primary operator, and
therefore can not correspond to a vacuum state.

\bigskip
\noindent{\bf Penrose limit \two}
\medskip

\noindent  In Penrose limit \two, the geometric action of $\Zori$,
$\vphi_3 \to \vphi_3 + \pi$ produces a reflection
of the transverse coordinate $w_3 \to -w_3$.
The geodesic lies in the orientifold plane $w_3=0$,
and the coincident D7-branes gives rise to an open string
sector in addition to the closed string sector.

This limit of the theory was considered in detail
by Berenstein et.~al.~\cite{Berenstein:2002b},
to which we refer the reader for a complete description
of the closed and open string sectors 
and the corresponding gauge theory operators.

\setcounter{equation}{0}
\section{$\cN=2$ $\su{N}$ with $2\protect\Yasymm\oplus4\protect\fund$}
\la{ssua}

The $\cN=2$ $\su{N}$ theory with two antisymmetric
and four fundamental hypermultiplets
contains an $\cN=2$ vector multiplet in the adjoint representation,
which in $\cN=1$ language
comprise a vector multiplet
and a chiral multiplet, ${\phi_a}^b$,
in the adjoint representation.
The theory also contains two antisymmetric $\cN=2$ hypermultiplets,
which in $\cN=1$ language
comprise a pair of chiral multiplets
$A_{iab}$ $(i=1,2)$ in the $\Yasymm$ representation
and a pair of chiral multiplets
$\tA_i^{ab}$ $(i=1,2)$ in the $\basymm$ representation.
Both $(A_1, A_2)$ and $(\tA_1,\tA_2)$ form SU$(2)_L$ doublets.
The conjugate fields $(A^{\dagger 1},A^{\dagger 2})$
transform in the conjugate SU$(2)_L$ representation,
but by lowering the indices with the $\epsilon$ tensor,
so that
$A^{\dagger 1}=A^\dagger_2$ and $A^{\dagger 2}=-A^\dagger_1$
we can form an SU$(2)_L$ doublet
$(A^\dagger_1, A^\dagger_2)$ from the conjugate fields
(and similarly for $(\tA^\dagger_1, \tA^\dagger_2)$).
Finally,
the theory contains four fundamental $\cN=2$ hypermultiplets,
which in $\cN=1$ language
comprise four chiral multiplets $Q^I_{a}$ in the $\fund$ representation
and four chiral multiplets $\tQ_I^{a}$ in the $\bfund$ representation.
The quantum numbers of the scalars in the above supermultiplets are
shown in the table below.

\vbox{
\begin{center}
\begin{tabular}{|c|c|c|c|c|c|c|c|}
\hline
CFT field & $\su{N}$ & $\JR$ & $J^3_R$ & $J^3_L$ &
$\Delta$ & $H_\one$ & $H_\two$
\\
\hline
& & & & & & & \\[-13pt]
\hline
& & & & & & & \\[-13pt]
$\phi$  	& adjoint 	& 1  &      0 &      0  &  1 & 0&1 \\
$\phi^\dagger$  & adjoint 	& $-1$ &      0 &      0  &  1 & 2&1 \\
$(A_1,\tA^\dagger_1)$& $\Yasymm$& 0 & $(\half,-\half)$&$\half$ &1
& (1,1)&(0,1)\\
$(A_2,\tA^\dagger_2)$& $\Yasymm$& 0 & $(\half,-\half)$& $-\half$
&1&(1,1)&(1,2)
\\[1pt]
$(\tA_1,A^\dag_1)$& $\basymm$& 0 & $(\half,-\half)$&$ \half$&1&(1,1)&(0,1)
\\[1pt]
$(\tA_2,A^\dag_2)$& $\basymm$& 0 & $(\half,-\half)$&$-\half$&1&(1,1)&(1,2) \\
$(Q^I,\tQ_I^\dagger )$ & $\fund$  	& 0 & $(\half,-\half)$&       0 &  1 & (1,1)&$(\half,\thrhalf)$ \\
$(\tQ_I,Q^{I\dagger})$   & $\bfund$  	& 0 & $(\half,-\half)$&       0 &  1 & (1,1)&$(\half,\thrhalf)$ \\[3pt]
\hline
\end{tabular}
\end{center}
\begin{center}
{\noindent\footnotesize{\bf Table \arabic{table}:\stepcounter{table}}
$\su{2}_R$ multiplets of scalar fields
in $\cN=2$ $\su{N}$\, with\,\,$2\,\Yasymm\oplus4\fund$\,.}
\end{center}
}
The F-term equations for the model are
\be
\label{FtermSU}
A_1 \tA_2 - A_2 \tA_1 + Q^I \tQ_I=0\,, \quad \phi A_i + A_i \phi^T  =0\,,
\;\; \tA_i \phi + \phi^T \tA_i=0\,, \quad \phi\, Q^I =0= \tQ_I \phi .
\ee

The $\cN=2$ $\su{N}$ gauge theory
just described arises as the worldvolume theory on a stack of 
coincident D3-branes in a flat type IIB background modded out by
the orientifold group
\be
\Gori = \Zorb \times \Zori,
\ee
where $\Zorb=\{\1,R_{6789}\}$
and $\Zori=\{\1,R_{45}\Omega (-1)^{F_L}\}$
act on the $\IC^3$ transverse to the D3-branes as
\be
z_3 \stackrel{\Zori}{\longrightarrow}-z_3\,,\qquad\qquad
(z_1,z_2)\stackrel{\Zorb}{\longrightarrow}(-z_1, -z_2)\,.
\ee
The fixed point set of $\Zori$ is the hyperplane $z_3=0$,
which corresponds to the position of an
O7-plane and four physical D7-branes,
while the fixed point set of $\Zorb$ is the
six-dimensional hyperplane $z_1=z_2=0$.

The fields of the $\su{N}$ theory discussed above
can be obtained via a projection from
those of the $\cN=4$ $\su{2N}$ theory\footnote{The projection really
gives a $\U(N)$ theory; however, since we are interested in the large-$N$ limit the $\U(1)$ factor is suppressed and can be ignored.}, whose three adjoint
$\cN=1$ chiral superfields are denoted $\Phi_i$.
The independent generators $\ga_\tha$ and $\ga_{\Om^\prime}$
of the orientifold group are realized as the $2N {\times} 2N$  matrices
\cite{Park:1998}
\begin{equation}
\label{PU2A}
\ga_{\tha}  = i \left( \begin{array}{cc}   \1_N & 0 \\
 					0  & -\1_N
	\end{array} \right ) \,, \qquad
\ga_{\Om'}  = \left( \begin{array}{cc} 0  & \1_N \\
 			-\1_N  & 0
\end{array} \right ) ,
\end{equation}
which induce the following projections on the fields of the gauge theory:
\bea
\Phi_{1,2} &= -\ga_{\tha} \Phi_{1,2} \ga_{\tha}^{-1}\,, & \qquad
\Phi_{3} = \ga_{\tha} \Phi_{3} \ga_{\tha}^{-1} \,,\non\\
\Phi_{1,2} &=  \ga_{\Om'} \Phi_{1,2}^T \ga_{\Om'}^{-1}\,, &\qquad
\Phi_{3} = - \ga_{\Om'} \Phi_{3}^T \ga_{\Om'}^{-1}\,,
\eea
resulting in $\Phi_i$'s of the form
\begin{equation}
\label{SU2APhi}
\Phi_1 = \left( \!\! \begin{array}{cc} 0 & A_1 \\
		-\tilde{A}^2 & 0 \end{array} \right)
       = \left( \begin{array}{cc} 0 & A_1 \\
		\tilde{A}_1 & 0 \end{array} \!\! \right)
\,, \quad
\Phi_2 =\left( \!\! \begin{array}{cc} 0 & A_2 \\
		\tilde{A}^1 & 0 \end{array} \!\!\right)
       =\left( \!\! \begin{array}{cc} 0 & A_2 \\
		\tilde{A}_2 & 0 \end{array} \!\! \right)
\,,\quad
\Phi_3 = \left( \!\! \begin{array}{cc} \phi & 0 \\ 0 &
		-\phi^T \end{array} \!\! \right).
\end{equation}

In the near-horizon (large-$N$) limit of the above orientifolded background
one obtains $\mathit{adS}_5{\times}S^5/[\Zorb{\times}\Zori]$, where
$\Zorb$ acts as $\vphi_1 \to \vphi_1 + \pi$, $\vphi_2 \to \vphi_2 + \pi$
and $\Zori$ acts as $\vphi_3 \to \vphi_3 + \pi$ on the coordinates of $S^5$.
String theory in this background is dual to the $\SU(N)$ gauge theory
discussed above (see ref.~\cite{Ennes:2000} for further details). Below we
discuss the Penrose limits for this correspondence.

\bigskip
\noindent{\bf Penrose limit \one}
\medskip

\noindent In the first Penrose limit,
the generator of $\Zori$ simply produces a translation
along the geodesic: $\vphi_3 \to \vphi_3 + \pi$
(together with the action of $\Omega (-1)^{F_L}$),
with no action on the transverse coordinates $w_1$ and $w_2$ (\ref{w1w2}).
There is no orientifold fixed plane
and therefore no open string sector in this limit,
but $\Zori$ projects out half of the discrete values of
$J_\one$ that would be allowed in the unprojected case.
The generator of $\Zorb$ acts solely on the transverse coordinates
as $w_1 \to -w_1$, $w_2 \to -w_2$.
The geodesic lies in the orbifold fixed plane,
resulting in a twisted sector of the string theory
in this pp-wave background.
First we describe the operators corresponding to the untwisted sector.

\medskip
\noindent{\it Untwisted sector}
\medskip

\noindent The identification of the gauge theory
operators corresponding to the states of the dual string theory in the
Penrose limit
is facilitated by
going over to the cover space, as in
refs.~\cite{Kim:2002,Mukhi:2002} (see also \cite{Douglas:1996}).
The ground state in the untwisted sector corresponds 
to $\tr[\Phi_3^{J_\one}]$.
As a result of the orientifold projection \eqs{SU2APhi},
this is only nonvanishing for even $J_\one$, yielding the
$H_\one=0$ operator
\be
\tr \left[  \phi^{2n} \right]\,.
\ee
The $H_\one=1$ gauge theory
operators $\tr[(D_{\mu} \Phi_3) \Phi_3^{2n+1}]$, or equivalently
using \eqs{SU2APhi},
\be
\tr \left[  (D_\mu \phi) \phi^{2n+1} \right]\,,
\ee
correspond to string states in which a single zero-momentum oscillator
$\adag^\mu$ in one of the directions of $\IR^4$
acts on the light-cone vacuum.

Since the zero-momentum oscillators $\adag^I$ in the $\tIR^4$
directions (\ref{w1w2}) are odd under $\Zorb$,
an even number of $\adag^I$'s must act on the light-cone vacuum.
The $H_\one =2$ gauge theory operators corresponding to the states in which
two such oscillators act on the vacuum
involve the insertion of two $H_\one=1$ fields
$\Ups_I = (\Phi_1,\Phi_1^\dagger, \Phi_2, \Phi^\dagger_2)$
into the ground state operator (the insertion of a single $H_\one=1$ operator
into the ground state operator
gives zero using \eqs{SU2APhi}):
\be
\sum_{m=0}^{J_{\one}}  \tr \left[ \Ups_I \Phi_3^m \Ups_K \Phi_3^{J_\one-m} \right],
\ee
where we average over the relative position of the
$\Ups$'s \cite{Berenstein:2002a,Kristjansen:2002,Constable:2002}.
The sum over $m$ vanishes unless $J_\one=2n$.
Using \eqs{SU2APhi}, together with
\be
\Phi_1^\dagger = \left( \begin{array}{cc} 0 & \tA^{\dagger 1} \\
		A^{\dagger 1} & 0 \end{array} \right)
               = \left( \begin{array}{cc} 0 & \tA^\dagger_2 \\
		A^\dagger_2 & 0 \end{array} \right)\,, \qquad
\Phi_2^\dagger =\left( \begin{array}{cc} 0 & \tA^{\dagger 2} \\
		 A^{\dagger 2} & 0 \end{array} \right)
		=\left( \begin{array}{cc} 0 & -\tA^\dagger_1 \\
		-A^\dagger_1 & 0 \end{array} \right) \,,\qquad
\ee
we obtain the set of ten operators
\bea
&{\ds \sum_m } (-)^m
\tr \left[ \tA_{(i}\phi^{m}  A_{j)} (\phi^T)^{2n-m} \right],\qquad
{\ds \sum_m} (-)^m
\tr \left[ A^\dag_{(i}\phi^{m}  \tA^\dag_{j)} (\phi^T)^{2n-m} \right],\non\\
&{\ds \sum_m} (-)^m
\tr \left[ A^\dag_{i} \phi^{m}  A_{j}       (\phi^T)^{2n-m}
        + \tA_{j}     \phi^{m} \tA^\dag_{i} (\phi^T)^{2n-m}  \right].
\ee
These correspond to the ten string theory states
\bea
\label{twoosc}
\adag^i \adag^j \vac\,, && \qquad\qquad\qquad
\badag^i \badag^j \vac \,, \ret
&&\badag^i \adag^j \vac \,,
\ee
where $\adag^i$ ($\badag^i$)
is the zero-momentum operator in the direction $w_i$ ($\bw_i$).

In limit \one{}  there are no open strings, since the D7-branes disappear.
The gauge theory explanation of this fact is that the candidate
open string vacuum $\tQ_I \phi^{2n} Q^K$ is ruled out since it is 
not a chiral primary operator, 
as can be seen from the fact that it vanishes 
by the F-term equation, $\phi \,Q^K =0$.

\medskip
\noindent{\it Twisted sector}
\medskip

\noindent The ground state in the twisted sector of the closed string theory
corresponds to the operator $\tr[\ga_{\tha}\Phi_3^{J_\one}]$.
The presence of the operator $\ga_\tha$ in the trace
requires $J_\one$ to be odd, giving
\be
\tr \left[  \phi^{2n+1} \right] .
\ee

The $H_\one=1$ operators corresponding to the string theory oscillators
in the $\IR^4$ directions of the
pp-wave metric are identified, in the twisted sector, with
$\tr[\ga_{\tha} (D_{\mu}\Phi_3) \Phi_3^{2n+2}]$, or
\be
\tr \left[ (D_\mu \phi) \phi^{2n+2} \right] .
\ee

The oscillators in the twisted sector are half-integer moded
\cite{Itzhaki:2002,Alishahiha:2002a,Kim:2002,Takayanagi:2002},
so there are no zero-momentum oscillators in this sector.
Hence the operators that would have corresponded to such 
oscillators acting on the twisted-sector vacuum
\be \label{notwist}
\sum_m  \tr \left[ \ga_\tha \Ups_I \Phi_3^m \Ups_J \Phi_3^{2n-m}
\right] ,
\ee
must not be chiral primary operators in the gauge theory.
Consider for example
\bea
&{\ds \sum_m} (-)^m
\tr \left[ \tA_{[i}\phi^{m}  A_{j]} (\phi^T)^{2n-m} \right].
\ee
By using the F-term equation $\tA_i \phi + \phi^T \tA_i=0$,
$\tA_i$ can be shifted through the $\phi^T$'s.
Then $A_{[j} \tA_{i]}$ can be replaced by $Q$'s,
using $A_1 \tA_2 - A_2 \tA_1 + Q^I \tQ_I=0$.
Finally, using $\phi\, Q^I =0$, the operator vanishes and
is therefore not a chiral primary operator.
The other operators in (\ref{notwist}) are similarly expected 
to be ruled out.

Although there are no zero-momentum oscillators in the twisted sector,
non-zero modes are present. 
The construction of the corresponding gauge theory operators  
proceeds in analogy with the discussions in 
refs.~\cite{Itzhaki:2002,Alishahiha:2002a,Kim:2002,Takayanagi:2002}.

\bigskip
\noindent{\bf Penrose limit \two}
\medskip

\noindent In limit \two, the geometric action of $\Zori$,
$\vphi_3 \to \vphi_3 + \pi$, produces a reflection
of the coordinate transverse to the O7-plane: $w_3 \to -w_3$,
 cf.~(\ref{w2w3}).
Thus the geodesic lies in the orientifold (O7) plane $w_3=0$,
and the coincident D7-branes give rise to an open-string sector.
The generator of $\Zorb$ produces a shift $\vphi_1 \to \vphi_1 + \pi$
halfway around the geodesic,
accompanied by the reflection $w_2 \to -w_2$.
Since the $\Zorb$ action has no fixed points,
there is no twisted sector in this Penrose limit,
however $\Zorb$ projects out half of the discrete values of
$J_\two$ that would be allowed in the unprojected case.
The pp-wave background in this limit is the same as the one
considered in ref.~\cite{Berenstein:2002b} except for the additional
action of $\Zorb$, which leads to different species of open string states,
not encountered in ref.~\cite{Berenstein:2002b}.

\medskip
\noindent{\it Closed string sector}
\medskip

\noindent The vacuum state of the closed string sector corresponds to the
$H_\two=0$ operator $\tr[\Phi_1^{J_\two}]$,
which, due to the $\Zorb$ projection, is only nonvanishing when $J_\two = 2n$,
\be \label{suBvac}
\tr[\Phi_1^{2n}] = \tr \left[ \left(A_1 \tA_1\right)^n \right].
\ee
(On the gauge theory side the constraint on $J_\two$
is required because we need an equal number
of $\Yasymm$ and $\basymm$ fields to form a gauge invariant operator.)
The absence of the twisted sector vacuum can been seen from the fact
that $\tr[\ga_\tha \Phi_1^m]=0$ for all values of $m$.

String states in which a zero-momentum oscillator
$\adag^\mu$ in one of the directions of $\IR^4$
acts on the light-cone vacuum
correspond to the $H_\two=1$ gauge theory
operator
\be
\tr[(D_{\mu}\Phi_1) \Phi_1^{2n+1}] =
\tr \left[ D_\mu (A_1 \tA_1) (A_1 \tA_1)^n \right].
\ee

String states in which a zero-momentum oscillator
$\adag$ ($\badag$) in the $w_2$ ($\bar{w}_2$) direction of $\tIR^4$
acts on the light-cone vacuum
corresponds to insertions of the $H_\two=1$
field $\Phi_2\Phi_1$ ($\Phi^\dagger_2\Phi_1$) into the vacuum state operator:
\bea
\tr \left[ \Phi_2  \Phi_1^{2n+1}  \right]
&= \tr \left[
(A_2 \tA_1 + A_1 \tA_2)  \left(A_1 \tA_1\right)^n  \right],\non\\
\tr \left[ \Phi_2^\dag \Phi_1^{2n+1}  \right]
&= \tr \left[
(\tA^\dagger_1 \tA_1 + A_1 A^\dagger_1)  \left(A_1 \tA_1\right)^n  \right].
\ee
Note that an insertion of $(A_1\tA_2 - A_2 \tA_1)$
can be replaced by $Q^I\tQ_I$ via the F-term equations,
and thus turns the would-be $H_\two=1$ closed string state
into the (subleading) open string vacuum (\ref{SU2Ameson}) 
(this appears to be related to the closed/open string interaction). 
Since the operator vanishes modulo subleading terms by using 
the F-term equations it is not protected and is thus ruled out.
There is no similar gauge theory argument which rules out the
insertion of the $H_\two =1$ operator
$A_1 A^\dagger_1 - \tA^\dagger_1 \tA_1$.
It would be interesting to
better understand why this operator is
ruled out\footnote{A similar comment applies to the $H_\two=1$ insertion,
$(D_\mu A_1) \tA_1 - A_1 (D_\mu \tA_1)$.}.

The zero-momentum oscillators in the $w_3$, $\bar{w}_3$
directions of $\tIR^4$
correspond to insertions of $\phi$ and $\phi^\dagger$
into the vacuum state (\ref{suBvac}).
These oscillators are odd under $\Zori$ and so
there should be no operators corresponding to single
oscillators acting
on the vacuum. This is automatic in the cover space language since
the gauge theory operators
$\tr [\Phi_3 \Phi_1^{2n}]$ and
 $\tr [\Phi_3^\dagger \Phi_1^{2n}]$ vanish identically.
An independent argument can also be constructed.
The   insertion of a single
$\phi$ is not possible since
$\tr[\phi (A_1 \tA_1)^n] =
\half \tr[(\phi A_1 + A_1 \phi^T)\tA_1(A_1\tA_1)^{(n-1)}]$, 
which vanishes by the F-term equations. 
Thus the operator is not a chiral primary
(nor is it a ``near chiral primary'') 
and is consequently ruled out.
We expect that the insertion of a single
$\phi^\dagger$ is also ruled out as implied by the cover space
result.

States with two oscillators in the $w_3$ direction
acting on the ground state are not projected out and correspond to
the insertion of a pair of $\Phi_3$'s
\be
\sum_{m=0}^{2n}  \tr \left[ \Phi_3 \Phi_1^m \Phi_3 \Phi_1^{2n-m} \right],
\ee
which gives
\be
\sum_{m~{\rm even}} \! \tr \left[ \phi
\left(A_1 \tA_1\right)^{\frac{m}{2}}  \phi
\left(A_1 \tA_1\right)^{n-\frac{m}{2}}\right]
- \!
\sum_{m~{\rm odd}} \! \tr \left[ \phi
\left(A_1 \tA_1\right)^{\frac{m{-}1}{2}}
\! \tA_1 \phi^T \tA_1 \left(A_1 \tA_1\right)^{n-\frac{m{+}1}{2}}\right].
\ee
States where one (or both) of the zero-momentum oscillators are in
the $\bar{w}_3$ direction are obtained from the above
expressions by replacing one (or both) of the $\Phi_3$'s by
$\Phi_3^\dagger$.

\medskip
\noindent{\it Open string sector}
\medskip

\noindent Since the geodesic lies in the orientifold (O7) fixed-plane
$w_3=0$ which is also the location of the D7-branes, there is an
open-string sector.
The gauge symmetry of the D7-branes/O7-plane, $\so{8}_F$,
is broken to $\su{4}_F {\times} \u{1}$ by the $\Zorb$ projection.
The ground state of the open string sector,
which transforms in the adjoint $\bf{28}$ of $\so{8}_F$,
decomposes into
\be
{\bf 28}
\to {\bf 15}_0 \oplus {\bf 1}_0 \oplus {\bf 6}_2 \oplus\bar{\bf 6}_{-2}\,.
\ee
The $\Zorb$ projection correlates the states
in the ${\bf 15}_0 \oplus {\bf 1}_0$ representation
with operators in odd-dimensional
representations of $\SU(2)_L$ \cite{Ennes:2000}.
These open string vacuum states should therefore correspond to
gauge theory operators
where the total number of $A$'s and $\tA$'s is even. There is a
natural candidate
\be
\la{SU2Ameson}
\tQ_{I}  (A_1 \tA_1)^{n} Q^{K} \,, \quad
\ee
which indeed belongs to the correct
(${\bf 15}_0\oplus{\bf 1}_0$) $\SU(4)_F{\times}\U(1)$ representation.
Furthermore, the operator \eqs{SU2Ameson} has $H_\two = 1$,
reflecting the non-vanishing zero point energy for
the open string sector~\cite{Berenstein:2002b}.

The $\Zorb$ projection correlates the states
in the ${\bf 6}_2 \oplus\bar{\bf 6}_{-2}$ representation
with operators in even-dimensional
representations of $\SU(2)_L$ \cite{Ennes:2000}.
These open string vacuum states should therefore correspond to
gauge theory operators where the total number of $A$'s and $\tA$'s is odd.
The natural candidates are
\be \label{SU2Adiquark}
Q^{[I} \tA_1 (A_1 \tA_1)^n Q^{K]} \,, \qquad \quad
\tQ_{[I} (A_1 \tA_1)^n A_1 \tQ_{K]} \,.
\ee
In these expressions, the flavor indices on the $Q$'s are antisymmetrized
because the operator sandwiched in between them is antisymmetric in
gauge indices, which shows that the operators (\ref{SU2Adiquark})  
belong to the correct
$\SU(4)$ representations. Furthermore, they have the correct
$\U(1)$ (quark number) charges.

Notice that in contrast to the model discussed in \cite{Berenstein:2002b}
there are two distinct types of open strings in this model.
The first, corresponding to \eqs{SU2Ameson}
are ``meson'' open strings with a quark at one end and an anti-quark
at the other, whereas the other, corresponding to \eqs{SU2Adiquark},
have quark number $\pm 2$, with quarks (or anti-quarks) at both ends.

The open string state in which a zero-momentum oscillator
in one of the $\IR^4$ directions act on the open string vacua
corresponds to the insertion of the operator $D_{\mu}(A_1\tA_1)$ into
the operators in \eqs{SU2Ameson} and \eqs{SU2Adiquark}.

Zero-momentum oscillators $\adag$, $\badag$
in the $w_2$, $\bw_2$ directions of $\tIR^4$
acting on the meson string vacuum (\ref{SU2Ameson})
correspond to the two $H_\two=2$ operators
\be
\sum_m \tQ_{I} (A_1 \tA_1)^{m} (A_1 \tA_2 + A_2\tA_1)  (A_1 \tA_1)^{n-m} Q^{K},
\ee
\be
\sum_m \tQ_{I} (A_1 \tA_1)^{m}
(\tA^\dagger_1 \tA_1 + A_1 A^\dagger_1)
(A_1 \tA_1)^{n-m} Q^{K}\,,
\ee
and analogous insertions into the  states \eqs{SU2Adiquark}.
The insertion of the other linear combination of $H_\two=1$ fields
$(A_1 \tA_2 -A_2\tA_1)$  results in splitting the
operator $\tQ \cdots Q$
into $\tQ \cdots Q \,\tQ \cdots Q$
by using the F-term equations.
The latter operator is a ``double-trace'' operator which is
subleading in the $1/N$-expansion.
Since the operator vanishes (modulo subleading terms) by using
the F-term equations it is not a chiral primary operator 
(nor is it a ``near chiral primary'') 
and is therefore ruled out.

Zero-momentum oscillators in the $w_3$, $\bw_3$ directions of $\tIR^4$
are absent from the open string spectrum \cite{Berenstein:2002b}, so the
corresponding operators on the gauge theory side should be absent as well.
This follows from the fact that the insertion of a $\phi$ 
(which has $H_\two=1$) can always be commuted to one of the endpoints 
by using the F-term equations,
where it vanishes using the $0= \tQ_I \phi$ F-term equation.
Hence, such an operator is not chiral primary, and is therefore not protected.
Similarly, insertions of $\phi^\dagger$ should not be allowed,
but this is more difficult to show.

The number of bosonic zero modes obtained above is in agreement
with the string theory result \cite{Berenstein:2002b}, where it was
shown that on the string theory side there are six bosonic and
four fermionic zero modes.

\setcounter{equation}{0}
\section{The $\cN=2$ $\SU{\times}\SU {\times}\cdots{\times} \SU{\times} \SU$ orientifold}
\label{sgoose1}
The $\su{N}+2\Yasymm\oplus4\fund$ theory studied in section~\ref{ssua}
is the first in an infinite series of similar
theories \cite{Park:1998}. 
The orientifold group for these theories
has the form $G = \Z^{\mathrm{orb}}_{2N_2}\times \Zori $, where
the pure orientifold part has the universal form
$\Zori = \{1,\Om(-1)^{F_L}R_{45}\} $,
and the orbifold part is
$\Z^{\mathrm{orb}}_{2N_2} = \{1,\tha,\ldots,\tha^{2N_2-1}\}$,
where $\tha$ acts on the 6789 directions as
$(z_1,z_2) \rar (e^{\pi i/N_2}z_1,e^{-\pi i/N_2}z_2)$.
For a particular choice of the orientifold projections acting on
the D3-branes, the gauge group becomes~\cite{Park:1998}
\be \label{SUgoose}
\SU(v_1) \times \cdots \times \SU(v_{N_2}) \,,
\ee
with the matter content (in $\cN=2$ language)
\bea \label{SUgoosematter}
 \Yasymm_{1} +\bigoplus_{j=2}^{N_2} (\fund_{j-1},\bfund_{j})
+\Yasymm_{N_2} + \bigoplus_{j=1}^{N_2} w_j \fund_{j} \,.
\eea
The $w_j$'s are non-negative integers
constrained by $\sum_{j=1}^{N_2} w_j = 4$.
The $v_j$'s obey constraints arising from the
vanishing of the beta-function(s) of the field theory \cite{Park:1998}.
The adS/CFT correspondence involves the large $N_1$ limit,
where $N_1 \equiv v_1$ denotes the rank of the first 
factor of the gauge group
(denoted by $N$ in the previous section).
It can be shown \cite{Naculich:2001b}
that, to leading order in $N_2/N_1$, the $v_j$'s corresponding 
to the different gauge group factors are
equal: $v_j = N_1 \left[ 1 + \cO(N_2/N_1)\right]$.
Thus for  $N_2\ll N_1$ the gauge group is essentially $\SU(N_1)^{N_2}$. 
In the following we will not assume that 
$N_2 \ll N_1$, although there are some 
motivations for making this assumption as will be  
discussed later in this section.
The quantum numbers of the scalar fields of 
the $\SU{\times}\cdots{\times}\SU$ gauge theory
are given in the table below.

\vbox{
\begin{center}
\begin{tabular}{|c|c|c|c|c|c|c|c|}
\hline
CFT field & Representation  &
$\JR$  &
$J_R^3$&
$J_L^3$&
$\Delta$  &
$H_\two$  \\
\hline
& & & & & & \\[-13pt]
\hline
& & & & & & \\[-13pt]
$\phi_r$ 	& $\mathrm{adjoint}_r$ 	& 1 &  0 &  0& 1 & 1 \\[1pt]
$\phi^\dag_r$ 	& $\mathrm{adjoint}_r$ 	&$-1$ &  0 &  0& 1 & 1 \\[1pt]
$(A_1,(\tA^1)^\dag)$ &$\Yasymm_1$&0 & $(\half,-\half)$& $\half$ & 1 & (0,1) \\[2pt]
$(\tA^1,(A_1)^\dag)$ &$\basymm_1$&0 & $(\half,-\half)$& $-\half$ & 1 & (1,2) \\[1pt]
$(A_2,(\tA^2)^\dag)$ &$\,\,\,\Yasymm_{N_2}$&0 & $(\half,-\half)$& $-\half$ & 1 & (1,2) \\[2pt]
$(\tA^2,(A_2)^\dag)$ &$\,\,\,\basymm_{N_2}$&0 & $(\half,-\half)$& $\half$ & 1 & (0,1) \\[2pt]
$(B_p,C_p^\dagger)$ &$(\bfund_{p-1},\fund_p)$&0 & $(\half,-\half)$& $\half$ & 1 & (0,1) \\[1pt]
$(C_p,B_p^\dagger)$ &$(\fund_{p-1},\bfund_p)$&0 & $(\half,-\half)$&$-\half$ & 1 & (1,2) \\[1pt]
$(Q_{r},\tQ_r^\dag)$ &$ \half w_r \fund_r$ & 0 &$(\half,-\half)$ & 0 & 1 & $(\half,\thrhalf)$\\[1pt]
$(\tQ_r,Q_{r}^\dag)$ & $\half w_r \bfund_r$ & 0 &$(\half,-\half)$ & 0 & 1 & $(\half,\thrhalf)$\\[3pt]
\hline
\end{tabular}
\end{center}
\begin{center}
{\noindent\footnotesize{\bf Table \arabic{table}:\stepcounter{table}}
$\su{2}_R$ multiplets of scalar fields
in the $\cN=2$ $\SU{\times}\cdots{\times}\SU$ orientifold theory.  }
\end{center}
}

In the table, $r$ takes the values $r=1,\ldots, N_2$,
whereas $p$ ranges from 2 to $N_2$.
The quantum number $J^3_L$ in the table above
denotes a $\U(1)_L$ quantum number. Note that there is
generically no $\SU(2)_L$ symmetry
(only when $N_2 =1$ is there such a symmetry).
The correspondence between our notation and that
of ref.~\cite{Mukhi:2002} is as follows:
$J^3_R \rar J^\prime$,
$J^3_L \rar N_2 J$.

We will now show that there is a limit of the 
$\SU{\times}\cdots{\times}\SU$ theory discussed above in which the
null direction $x^-$ becomes compact,
analogous to the limit discovered in refs.~\cite{Mukhi:2002,Alishahiha:2002b}.
Although the gauge group of the model is very similar to that
of the unorientifolded model considered in 
refs.~\cite{Mukhi:2002,Alishahiha:2002b},
the matter content differs;
in particular, there are no bifundamental fields
connecting the last and first factors of the gauge group.

We consider only Penrose limit \two{},
since it is in this limit that the compact null direction appears.
Using the results of section \ref{spen}, 
we find that the orbifold identifies  
$(\vphi_1, w_2)$ with $(\vphi_1 + (\pi/N_2), \e^{-i \pi/N_2} w_2)$.
This identification along the geodesic translates into
$x^+ \sim x^+ + (\pi/2N_2)$ and $x^- \sim x^- + (\pi R^2/2N_2)$.
As observed in \cite{Mukhi:2002},
if one take a scaling limit in which
$2\pi R_- \equiv \pi R^2/2N_2 \sim (g_{\mathrm{s}} N_1/N_2)^{1/2}\al'$
stays finite,
the $x^-$ direction becomes compact with radius $R_-$.
Since we obtain a background with a compact
null direction the string theory can be quantized using the
discrete-light-cone-quantization (DLCQ) scheme, familiar from matrix
theory \cite{Susskind:1997}.

\medskip
\noindent{\it Closed string sector}
\medskip

\noindent Since there is no representation connecting
the last and first factor of the product gauge group,
a gauge invariant operator cannot be constructed
{}from a ``loop'' of $H_\two=0$ bifundamental fields $B_p$,
as in ref.~\cite{Mukhi:2002}.
However, we may introduce the two $H_\two = 0$ operators
\bea \label{halfcs}
\begin{array}{rcl} 
\mathcal{A}_{a\bb} &=& (A_1 B_2 \cdots B_{N_2})_{a\bb}   \\
\tilde{\mathcal{A}}^{a \bb} &=&
(B_2 \cdots B_{N_2} \tA^{2})^{a \bb} 
\end{array} \qquad \quad
\left\{ \begin{array}{l}
a = 1, \cdots, v_1 \\
\bb = 1, \cdots, v_{N_2} 
\end{array} \right.
\eea
{}From these, a gauge invariant operator
\be
\la{moosevac}
\tr[(\mathcal{A}\tilde{\mathcal{A}}^{T})^k]
\ee
can be constructed which may be identified with
vacuum of the closed string sector.
By the reasoning in ref.~\cite{Mukhi:2002},
$k$ is the momentum labeling the different DLCQ sectors.

In ref.~\cite{Mukhi:2002}, the ground state corresponded to
a ``loop'' of bifundamental fields, wrapping $k$  times
around the quiver/moose diagram.
It was suggested that this was related to the T-dual
type IIA picture of the gauge theory.
In a similar way, the two operators in (\ref{halfcs}) 
can be identified with two half-circles of
a quiver/moose diagram which are identified by the orientifold.
This is reminiscent of the type IIA picture of the orientifold model,
which contains two O$6^-$ planes located at diametrically
opposite points of the elliptic circle,
which reflect the circle onto itself \cite{Park:1998}.
It should be possible to connect these two pictures 
by applying a T-duality as in ref.~\cite{Mukhi:2002}.
(T-dualities of the pp-wave background have also been discussed
in ref.~\cite{Cvetic:2002}).

The operators corresponding to the string theory states
may also be obtained using a cover space approach, as in
ref.~\cite{Mukhi:2002} and in sec.~\ref{ssua}.
Beginning with matrices $\Phi_i$ in the adjoint representation
of an $\SU(2 N_v)$ theory,
where $N_v = \sum_r v_r$, and projecting to 
the $\SU(v_1){\times}\cdots{\times}\SU(v_{N_2})$ theory using 
a generalization \cite{Park:1998} of eqs.~\eqs{PU2A},  we obtain
\be \label{goosePhis}
\Phi_1 = \left( \!\! \begin{array}{cc} {\bf B}^T & {\bf A_1}  \\
		-{\bf \tilde{A}^2} & {\bf B} \end{array} \right)
\,, \quad
\Phi_2 =\left( \!\! \begin{array}{cc} {\bf C}  & {\bf A_2} \\
		{\bf \tilde{A}^1} & {\bf C}^T \end{array} \!\!\right)
\,,\quad
\Phi_3 = \left( \!\! \begin{array}{cc} {\bf \phi} & 0 \\ 0 &
		-{\bf \phi}^T \end{array} \!\! \right),
\ee
where 
$ \phi = \diag(\phi_1, \phi_2 , \ldots, \phi_{N_2}) $ and
\bea \label{ABC}
{\bf A_1} &=& \diag(A_1,0,\ldots,0)\,, \qquad\qquad 
{\bf \tA^1} = \diag(\tA^1,0,\ldots,0)\,,\non \\
{\bf A_2} &=& \diag(0,\ldots,0,A_2)\,, \qquad \qquad
{\bf \tA^2} = \diag(0,\ldots,0,\tA^2)\,, \non
\eea
\bea
{\bf B} &=& \left( \begin{array}{cccc}    0 & B_2 &\cdots   & 0 \\
                                0 &   0 & \ddots  & \vdots \\
                           \vdots &     & \ddots  & B_{N_2} \\
                                0 &   0 & \cdots  & 0    \\ 
\end{array} \right), \qquad 
{\bf C} = \left( \begin{array}{cccc}        0 & 0 &\cdots  & 0 \\
                                C_2 & 0 &        & 0 \\
                                 \vdots & \ddots  & \ddots  & \vdots \\
                                  0 & \cdots & C_{N_2}  & 0 \\ 
\end{array} \right).
\ee
In terms of the $\Phi_i$'s, the ground state in the closed string sector with lightcone momentum $k$ can 
be written as
\be
\la{moosecover}  \tr (\Phi_1^{2N_2 k})\,,
\ee
which precisely coincides with eq.~\eqs{moosevac}.
Moreover, operators corresponding to string theory states
with zero-momentum oscillators acting on the vacuum can be
obtained via insertions of the $H_B=1$ fields
$(D_\mu \Phi_1) \Phi_1$,
$\Phi_2 \Phi_1$,
$\Phi_2^\dag \Phi_1$,
$\Phi_3 \Phi_1$,
and $\Phi_3^\dag\Phi_1$
into the vacuum operator, eq.~\eqs{moosecover}.

Since we have a compact direction, winding modes are present in the 
string theory. These and the non-zero-momentum states can be 
constructed as in ref.~\cite{Mukhi:2002}. 
As an example, a state with a single non-zero-momentum 
oscillator $a_n^\dag$ in the $w_3$ direction 
acting on the vacuum can be obtained through an insertion  
of $\Phi_3  V^n$ into (\ref{moosecover}). 
(Due to the $\Zori$ projection, this vanishes if $n=0$.)
Here $V$ is the diagonal matrix of phases 
$\sqrt{\om}~{\rm diag} (
\1_{v_1}, \om \1_{v_2}, \ldots, \om^{N_2-1} \1_{v_{N_2}}; 
\om^{2N_2-1} \1_{v_1}, \om^{2N_2-2} \1_{v_2},\ldots,\om^{N_2} \1_{v_{N_2}})$ 
with $\om= \exp (\pi i/N_2)$ \cite{Park:1998}.
As explained in \cite{Mukhi:2002}, such a state has winding number $n$.
More generally, if the sum of the oscillator mode numbers does not vanish,
then the state has non-zero winding number.

\medskip
\noindent{\it Open string sector}
\medskip

\noindent Since the geodesic in Penrose limit \two~lies in
the orientifold (O7) fixed-plane $w_3=0$,
which is coincident with the D7-branes, 
there is an open-string sector,
in contrast to the model studied in ref.~\cite{Mukhi:2002}.
The model
(\ref{SUgoose}), (\ref{SUgoosematter})
actually comprises several different models,
due to the fact that there are several possible choices for the 
(non-negative) $w_r$'s satisfying $\sum_r w_r =4$. The different 
models are distinguished by the fact that the fundamentals 
belong to different 
factors of the gauge group.
To be able to treat all the different cases
on the same footing we will use the following notation.
Assuming that $Q^I$ (or $\tQ_I$) for a particular $I$ 
belongs to the $r$th factor of the gauge group,
we can define
\bea
\cQ^{I}_{\ba} &=& (Q^I B_{r+1} \cdots B_{N_2})_{\ba}\,, \non \\
\tcQ_{I}^{a} &=&  (B_{2} \cdots B_{r} \tQ_I )^{a}\,.
\ee
{}From these, one can construct operators corresponding
to the open string vacua, 
in analogy with (\ref{SU2Ameson}), (\ref{SU2Adiquark})
\bea
\la{openstringop}
\cQ^I \tcA^T \left( \cA \tcA^T  \right)^{(k-1)} A_1 \tcQ_K, \qquad
\cQ^I \tA^2 \left( \cA^T \tcA \right)^{k} (\cQ^K)^T, \qquad
(\tcQ_I)^T \left( \cA^T \tcA \right)^{k} A_1 \tcQ_K.
\eea
As above, $k$ is the non-negative integer lightcone momentum quantum
number labeling the different DLCQ vacua. 
There are two kinds of open strings, as in section~\ref{ssua}.
(The operators \eqs{openstringop} reduce to those in the previous
section when $N_2=1$.)

Excited states in the open string sector 
are constructed by inserting $H_\two =1$ operators,
as in section~\ref{ssua}. 
Since there is no level matching condition in the open string sector, 
a single non-zero-momentum oscillator acting 
on the vacuum state is allowed. 

Finally, we will discuss the implications of making the additional 
assumption $N_2\ll N_1$.
For the original model \cite{Berenstein:2002a} the 't Hooft coupling
is $\lambda'=g_{\mathrm{YM}}^2 N/J^2$. 
It was recently argued~\cite{Constable:2002} that the
effective genus counting parameter in the Yang-Mills theory is
$g_2^2 = {J^4}/{N^2}$, and the effective coupling between a wide
class of excited states is $g_2\sqrt{\lambda'}$ 
(see ref.~\cite{Constable:2002} for further details; 
see also \cite{Gopakumar:2002} for some refinements). 
Translating these expressions to our case 
one finds that both the expansion parameters are small provided 
$N_2\ll N_1$, thus from this perspective it is natural to make the 
additional assumption $N_2\ll N_1$. 
Note that this assumption corresponds to a large radius approximation 
for the compact circle.

\setcounter{equation}{0}
\section{$\cN=2$ $\sp{2N}{\times}\sp{2N}$ with
$(\protect\fund,\protect\fund)\oplus
2(1,\protect\fund)\oplus 2(\protect\fund,1)$ }
\la{ssp1}

The $\cN=2$ $\sp{2N}{\times} \sp{2N}$ theory with
an $\cN=2$ hypermultiplet in the bifundamental representation $(\fund,\fund)$
and two $\cN=2$ hypermultiplets in the fundamental representation
of each factor of the gauge group contains (in $\cN=1$ language)
two chiral multiplets ${\phi_{1a}}^b$ and  ${\phi_{2\ba}}^\bb$
in the adjoint representation, one for each $\sp{2N}$ factor.
In matrix notation, these fields satisfy
\be
\la{Spconstraint}
\phi_1 = J_1 \phi_1^T J_1 \,,\qquad \phi_2 = J_2 \phi_2^T J_2\,,
\ee
where $J^{ab}= J_1$  and $J_{ab} = J_1^{-1} = -J_1$
is the symplectic unit of the first $\sp{2N}$ factor, used to
raise and lower indices
(and similarly for $J^{\ba\bb} = J_2$).
The theory also contains two chiral multiplets in the
$(\fund,\fund)$ representation;
for convenience, in writing operators in matrix notation,
we represent these with two chiral multiplets
${A_{ia}}^\bb$ $(i=1,2)$ in the $(\fund,\bfund)$
representation of $\Sp(2N){\times}\Sp(2N)$ and
two chiral multiplets ${B_{i\ba}}^b$ $(i=1,2)$
in the $(\bfund,\fund)$ representation,
with the constraint
\be
\la{ABconstraint}
B_i = -J_2 A_i^T J_1\,.
\ee
In this paper, both $(A_1, A_2)$ and $(B_1, B_2)$
are in the $\fund$ of SU$(2)_L$.
(The indices may be raised with the $\epsilon$ tensor,
so that $(B^1, B^2) \equiv (B_2, -B_1)$ is in the
$\bfund$ of SU$(2)_L$.
In ref.~\cite{Naculich:2002a} and many other papers, $(B^1, B^2)$
is written as $(B_1, B_2)$.)

Finally, the theory also contains four chiral multiplets
in the $(\fund,1)$ and $(1,\fund)$ representations,
denoted by $Q^I_{1a}$ and $Q^I_{2\ba}$ ($I=1,\ldots,4)$, respectively.
For convenience, we also introduce fields
${\tQ_{1I }^{ a}} $ and ${\tQ_{2I }^{\ba}} $
in the $(\bfund,1)$ and $(1,\bfund)$ representations,
related to the above fields by \cite{Naculich:2002a}
\be
\tQ_{1I}^T =  - g_{IJ } J_1  Q_1^{J} , \qquad
\tQ_{2I}^T=   - g_{IJ}  J_2  Q_2^{J} ,
\ee
where $g_{IJ}$ is the metric for either of the two factors of the
$\SO(4){\times}\SO(4)$ flavor symmetry group.
The quantum numbers of the scalar fields
are displayed in the table below.

\vbox{
\begin{center}
\begin{tabular}{|c|c|c|c|c|c|c|c|}
\hline
CFT field &
$\sp{2N}{\times} \sp{2N}$  &
$\JR$  &
$J_R^3$&
$J_L^3$&
$\Delta$  &
$H_\one$ &
$H_\two$ \\
\hline
& & & & & & & \\[-13pt]
\hline
& & & & & & & \\[-13pt]
$\phi_1$ 	& (adjoint,1) 	& 1 &      0 &        0& 1 & 0 & 1\\
$\phi_1^\dagger$& (adjoint,1) 	&$-1$ &      0 &        0& 1 & 2 & 1\\
$\phi_2$ 	& (1, adjoint) 	& 1 &      0 &        0& 1 & 0 & 1\\
$\phi_2^\dagger$& (1, adjoint) 	&$-1$ &      0 &        0& 1 & 2 & 1\\
$(A_1,B_1^\dagger)$&$(\fund,\bfund)$&0 & $(\half,-\half)$& $\half$ & 1 & (1,1) & (0,1)\\
$(A_2,B_2^\dagger)$&$(\fund,\bfund)$&0 & $(\half,-\half)$& $-\half$ & 1 & (1,1) & (1,2)\\
$(B_1,A_1^\dagger)$&$(\bfund,\fund)$&0 &$(\half,-\half)$ &$ \half$ & 1 & (1,1) & (0,1)\\
$(B_2,A_2^\dagger)$&$(\bfund,\fund)$&0 &$(\half,-\half)$ &$-\half$ & 1 & (1,1) & (1,2)\\
$(Q_1,\tQ_1^\dagger)$  &$ (\fund, 1)$   & 0 &$(\half,-\half)$ & 0 & 1 & (1,1)  & $(\half,\thrhalf)$\\[2pt]
$(\tQ_{1},Q_1^\dagger)$ &$ (\bfund, 1)$ & 0 &$(\half,-\half)$ & 0 & 1 & (1,1)  & $(\half,\thrhalf)$\\[2pt]
$(Q_2,\tQ_2^\dagger)$  &$ (1, \fund)$   & 0 &$(\half,-\half)$ & 0 & 1 & (1,1)  & $(\half,\thrhalf)$\\[3pt]
$(\tQ_{2},Q_2^\dagger)$&$ (1, \bfund)$  & 0 &$(\half,-\half)$ & 0 & 1 & (1,1)  & $(\half,\thrhalf)$\\[3pt]
\hline
\end{tabular}
\end{center}
\begin{center}
{\noindent\footnotesize{\bf Table \arabic{table}:\stepcounter{table}}
$\su{2}_R$ multiplets of scalar fields
in $\cN=2$ $\sp{2N} {\times} \sp{2N}$ with
$(\fund,\fund)\oplus 2(1,\fund)\oplus 2(\fund,1)$.}
\end{center}
}
The superpotential for this theory is
\be
\label{N=2super}
\mathcal{W}_{\cN=2} =
\tr \left[ \phi_1 (A_1 B^1 + A_2 B^2) - \phi_2 (B^1 A_1 + B^2 A_2) \right]
+ \tQ_{1I} \phi_1 Q_1^{I}
+ \tQ_{2I} \phi_2 Q_2^{I}\,,
\ee
giving rise to the F-term equations
\bea
\label{FtermSp}
&A_1 B_2 - A_2 B_1 + Q_{1}^I  \tQ_{1I} =0\,, \qquad
A_i \phi_2 - \phi_1 A_i=0\,, \qquad
\phi_1 Q^I_1 =0\,,      \non\\
&B_1 A_2 - B_2 A_1 + Q_{2}^I  \tQ_{2I} = 0\,,\qquad
B_i \phi_1 - \phi_2 B_i=0 \,,\qquad
\phi_2 Q^I_2=0 \,.
\ee

The $\cN=2$ $\sp{2N}{\times}\sp{2N}$ gauge theory just discussed
arises as the theory on a stack of D3-branes 
in the same IIB background that gave rise to the
$\su{N}$ with $2 \Yasymm \oplus 4 \fund$ theory discussed in
section~\ref{ssua},
namely, flat space modded out by
\be
\Gori = \Zorb \times \Zori\,.
\ee
As in that case, the fixed point set of $\Zori$ is the hyperplane $z_3=0$,
which corresponds to the position of the
O7-plane and the D7-branes, while the fixed point set of $\Zorb$ is the
six-dimensional hyperplane $z_1=z_2=0$.

The fields of the $\cN=2$ $\Sp(2N){\times}\Sp(2N)$ theory can
be obtained via a projection 
(different from the one in section \ref{ssua}) 
{}from the $\cN=4$ $\SU(4N)$ gauge theory.
The independent generators of the orientifold group
are realized as the $4N {\times} 4N$  matrices
\cite{Park:1998}
\begin{equation}
\label{PUSpSp}
\ga_{\tha}  = \left( \begin{array}{cc}   \1_{2N} & 0 \\
                    0  & -\1_{2N} \end{array} \right ) \,, \qquad
\ga_{\Om'}  = \left( \begin{array}{cc} J  & 0 \\
 			0  & J \end{array} \right ) .
\end{equation}
The orbifold projection
\be
\Phi_{1,2} = -\ga_{\tha} \Phi_{1,2} \ga_{\tha}^{-1}, \qquad
\Phi_{3} = \ga_{\tha} \Phi_{3} \ga_{\tha}^{-1},
\ee
restricts the adjoint fields, $\Phi_i$, of the $\cN=4$ theory
to be of the form
\begin{equation}
\label{SpSpPhi}
\Phi_1 = \left( \!\! \begin{array}{cc} 0 & A_1 \\
		-B^2 & 0 \end{array} \!\!\right)
       = \left( \!\! \begin{array}{cc} 0 & A_1 \\
		 B_1 & 0 \end{array} \right)\,, \quad
\Phi_2 =\left( \!\! \begin{array}{cc} 0 & A_2 \\
			B^1 & 0 \end{array} \!\! \right)
       =\left( \!\! \begin{array}{cc} 0 & A_2 \\
			B_2 & 0 \end{array} \!\! \right) \,,\quad
\Phi_3 = \left( \!\! \begin{array}{cc} \phi_1 & 0 \\ 0 &
		 \phi_2 \end{array} \!\!\right).
\end{equation}
The orientifold projection
\be
\la{oriconstraint}
\Phi_{1,2} =  \ga_{\Om'} \Phi_{1,2}^T \ga_{\Om'}^{-1}, \qquad
\Phi_{3} = - \ga_{\Om'} \Phi_{3}^T \ga_{\Om'}^{-1},
\eea
is equivalent to the constraints \eqs{Spconstraint} and \eqs{ABconstraint}.

The action of $\Gori$
on the coordinates of $\mathit{adS}_5{\times}S^5$ is as follows.
$\Zorb$ acts as $\vphi_1 \to \vphi_1 + \pi$, $\vphi_2 \to \vphi_2 + \pi$
and $\Zori$ acts as $\vphi_3 \to \vphi_3 + \pi$.
String theory on this background is dual to the above $\Sp{\times}\Sp$
gauge theory \cite{Gukov:1998,Ennes:2000}.
Below we discuss the pp-wave limits of this correspondence.

\bigskip
\noindent{\bf Penrose limit \one}
\medskip

\noindent In the first Penrose limit,
the generator of $\Zori$ simply produces a translation
along the geodesic: $\vphi_3 \to \vphi_3 + \pi$
(together with the action of $\Omega (-1)^{F_L}$)
with no action on the transverse coordinates $w_1$ and $w_2$ (\ref{w1w2}).
There is no orientifold fixed plane
and therefore no open string sector in this Penrose limit,
but $\Zori$ projects out half of the discrete values of
$J_\one$ that would be allowed in the unprojected case.
The generator of $\Zorb$ acts solely on the transverse coordinates
as $w_1 \to -w_1$, $w_2 \to -w_2$.
The geodesic lies in the orbifold fixed plane,
resulting in a twisted sector of the string theory
in this pp-wave background.

\medskip
\noindent{\it Untwisted and twisted sectors}
\medskip

\noindent As in the case of the $\SU(N) + 2\Yasymm \oplus 4\fund$
theory in section~\ref{ssua},
the operators in the untwisted and twisted sectors are easily
obtained through projections of the cover space fields.

The ground state in the untwisted sector corresponds to
the operator $\tr[\Phi_3^{J_\one}]$ whereas the twisted sector ground state
corresponds to the operator $\tr[\ga_{\tha}\Phi_3^{J_\one}]$.
The $\Zori$ projection \eqs{oriconstraint} implies
$\tr[\ga_\tha^q \Phi_3^{J_\one}]
= (-)^{J_\one} \tr[\ga_\tha^q (\Phi_3^T)^{J_\one}]
= (-)^{J_\one} \tr[\ga_\tha^q \Phi_3^{J_\one}] $,
which vanishes unless $J_\one = 2n $ for both
the untwisted and twisted sectors.
Thus, the $H_\one=0$ operators
\be
\tr \left[   \phi_1^{2n} \pm \phi_2^{2n} \right],
\ee
correspond to the vacuum states in the
untwisted ($+$) and twisted ($-$) sectors of the string theory.

The $H_\one=1$ gauge theory operators
$\tr[(D_{\mu}\Phi_3) \Phi_3^{2n+1}]$ and
$\tr[\ga_{\tha} (D_{\mu}\Phi_3) \Phi_3^{2n+1}]$
or, explicitly,
\be
\tr \left[ \left(D_\mu \phi_1\right) \phi_1^{2n+1} \pm
           \left(D_\mu \phi_2\right) \phi_2^{2n+1} \right],
\ee
correspond to string states in which a single zero-momentum oscillator
$\adag^\mu$ in one of the directions of $\IR^4$
acts on the light-cone vacuum in the untwisted ($+$) and twisted
($-$) sectors respectively.

Since the zero-momentum oscillators $\adag^I$ in the transverse $\tIR^4$
directions (\ref{w1w2}) 
are odd under $\Zorb$,
only pairs of such oscillators can act on the light-cone vacuum.
The $H_\one=2$ gauge
theory operators corresponding to the states involving two
oscillators in the untwisted sector
involve the insertion of a pair of $H_\one=1$ fields
$\Ups_I = (\Phi_1,\Phi_1^\dagger, \Phi_2, \Phi^\dagger_2)$
into the ground state operator (the insertion of a single $J_\one=1$ operator
into the ground state operator gives zero using \eqs{SpSpPhi}):
\be
\la{SpSpuntwist}
\sum_{m=0}^{J_\one}
\tr \left[ \Ups_I \Phi_3^m \Ups_K \Phi_3^{J_\one-m} \right],
\ee
where we average over the relative position of the $\Ups$'s
\cite{Berenstein:2002a,Kristjansen:2002,Constable:2002}.
Eq.~\eqs{SpSpuntwist} vanishes unless $J_\one$ is even.
Using \eqs{SpSpPhi}, together with
\be
\Phi_1^\dagger = \left( \begin{array}{cc} 0 & B^{\dagger 1}\\
		A^{\dagger 1} & 0 \end{array} \right)
               = \left( \begin{array}{cc} 0 &  B^\dagger_2 \\
		A^\dagger_2 & 0 \end{array} \right)\,, \qquad
\Phi_2^\dagger =\left( \begin{array}{cc} 0 & B^{\dagger 2}\\
		 A^{\dagger 2} & 0 \end{array} \right)
		=\left( \begin{array}{cc} 0 &  -B^\dagger_1 \\
		-A^\dagger_1 & 0 \end{array} \right) \,,\qquad
\ee
we obtain the set of ten $H_\one=2$ operators
\bea
&{\ds \sum_m}
\tr \left[ A_{(i}  \phi_2^{m}  B_{j)} \phi_1^{2n-m} \right],
\qquad {\ds \sum_m}
\tr \left[ B^\dag_{(i} \phi_2^{m} A^\dag_{j)} \phi_1^{2n-m} \right],
\non\\
&{\ds \sum_m}
\tr \left[ A_i \phi_2^{m} A^\dag_j \phi_1^{2n-m}
         + B^\dag_j \phi_2^{m} B_i \phi_1^{2n-m} \right],
\ee
in agreement with string theory expectations, cf.~\eqs{twoosc}.

The oscillators in the directions $w_1$ and $w_2$
are half-integer moded in the twisted sector
\cite{Itzhaki:2002,Alishahiha:2002a,Kim:2002,Takayanagi:2002}
so there exist no zero-momentum oscillators in this sector.
The operators that would have corresponded to such 
oscillators acting on the twisted-sector vacuum
\be
\la{SpSptwist}
\sum_m  \tr \left[\ga_\tha \Ups_I \Phi_3^m \Ups_K \Phi_3^{J_\one-m} \right],
\ee
must therefore not be chiral primary operators in the gauge theory.
Consider for example
\bea
\sum_m \tr \left[ A_{[i}  \phi_2^{m}  B_{j]} \phi_1^{J_\one-m} \right],
\ee
Using the F-term equation $ A_i \phi_2 = \phi_1 A_i$,
$A_i$ can be shifted through $(\phi_2)^m$.
Then $A_{[i} B_{j]}$ can be replaced by $Q_1$'s,
using  $A_1 B_2 - A_2 B_1 + Q_{1}^I  Q_{1I} =0$.
Finally, using $\phi_1 Q^I_1 =0$, the operator vanishes
and is therefore not a chiral primary.
The other operators in (\ref{SpSptwist}) are
similarly expected to be ruled out.

Operators corresponding to non-zero-momentum oscillators 
can be constructed in analogy with the constructions 
in refs.~\cite{Alishahiha:2002a,Kim:2002,Takayanagi:2002}.

\bigskip
\noindent{\bf Penrose limit \two}
\medskip

\noindent In limit \two, the geometric action of $\Zori$,
$\vphi_3 \to \vphi_3 + \pi$, produces a reflection
of the transverse coordinate $w_3 \to -w_3$.
The geodesic lies in the orientifold (O7) plane $w_3=0$,
and the coincident D7-branes give rise to an open-string sector.
The generator of $\Zorb$ produces a shift $\vphi_1 \to \vphi_1 + \pi$
halfway around the geodesic,
accompanied by a reflection of the
coordinates transverse to the O7-plane: $w_2 \to -w_2$.
Since the $\Zorb$ action has no fixed points,
there is no twisted sector in this Penrose limit,
however $\Zorb$ projects out half of the discrete values of
$J_\two$ that would be allowed in the unprojected case.

\medskip
\noindent{\it Closed string sector}
\medskip

\noindent The vacuum state of the closed string sector corresponds to the
$H_\two=0$ operator $\tr[\Phi_1^{J_\two}]$.
Due to the $\Zorb$ projection on $\Phi_1$,
this is only nonvanishing when $J_\two = 2n$,
\be
\tr[\Phi_1^{2n}] = \tr \left[ \left(A_1 B_1\right)^n \right].
\ee

String states in which a zero-momentum oscillator
$\adag^\mu$ in one of the directions of $\IR^4$
acts on the light-cone vacuum
correspond to the $H_\two=1$ gauge theory operators
\be
\tr[(D_{\mu}\Phi_1)\Phi^{2n+1}] =
\tr \left[ D_\mu (A_1 B_1) (A_1 B_1)^n \right].
\ee

String states in which a zero-momentum oscillator
$\adag$, $\badag$ in the $w_2$, $\bw_2$ directions of $\tIR^4$
acts on the light-cone vacuum
corresponds to insertions of the $H_\two=1$
fields $\Phi_2\Phi_1$, $\Phi_2^\dagger\Phi_1$ 
into the vacuum state operator:
\be
\tr \left[ \Phi_2 \Phi_1^{2n+1}  \right]
&= \tr \left[
(A_2 B_1 + A_1 B_2)  \left(A_1 B_1\right)^n  \right],\non\\
\tr \left[ \Phi_2^\dag \Phi_1^{2n+1}  \right]
&= \tr \left[
(B^\dagger_1 B_1 + A_1 A^\dagger_1)  \left(A_1 B_1\right)^n  \right].
\ee
The linear combination $A_2 B_1 - A_1 B_2$
is ruled out since it is not a chiral primary;
see section \ref{ssua} for further details.

The zero-momentum oscillators in the $w_3$ ($\bw_3$) directions 
of $\tIR^4$ are odd under $\Zori$,
so there are no states corresponding to single oscillators acting
on the vacuum.
Correspondingly, the gauge theory operator 
$\tr [\Phi_3 \Phi_1^{J_\two}]$, resp. $\tr [\Phi_3^\dagger \Phi_1^{J_\two}]$,
vanishes, 
as can be seen by taking the transpose of
$[\phi_1 (A_1 B_1)^n]$, resp. $[\phi^\dagger_1 (A_1 B_1)^n]$, 
and using \eqs{Spconstraint} and \eqs{ABconstraint}.

\medskip
\noindent{\it Open string sector}
\medskip

\noindent Since the geodesic lies in the orientifold (O7) fixed-plane
$w_3=0$ which is also the location of the D7-branes,
there is an open-string sector.
The gauge symmetry of the D7-branes in the presence of the O7-plane,
$\so{8}_F$,
is broken to $\so{4} {\times} \so{4}$ by the $\Zorb$ projection.
The ground state of the open string sector,
which transforms in the adjoint $\bf{28}$ of $\so{8}_F$,
decomposes into
\be
{\bf 28}
\to {\bf (6,1)} \oplus {\bf (1,6)} \oplus {\bf (4,4)}\,.
\ee
The $\Zorb$ projection correlates the states
in the ${\bf (6,1)} \oplus {\bf (1,6)}$
representation with operators in odd-dimensional
representations of $\SU(2)_L$ \cite{Ennes:2000}.
These open string vacuum states therefore correspond to
gauge theory operators with an even number of $A$'s and $B$'s.
The natural candidates are
\be
\la{SpSpsame}
Q_1^{[I} J_1 (A_1 B_1)^{n} Q_1^{K]} \,, \qquad \quad
Q_2^{[I} J_2 (B_1 A_1)^{n} Q_2^{K]}\,.
\ee
These operators have $H_\two = 1$, reflecting the non-vanishing
zero point energy for the open string sector~\cite{Berenstein:2002b}.
The antisymmetry of the flavor indices $I$, $K$
is enforced by the antisymmetry of the matrices
$J_1(A_1 B_1)^n$ and $J_2(B_1 A_1)^n$ (which follows from
eq.~\eqs{ABconstraint}) and guarantees that the operators \eqs{SpSpsame}
belong to the correct representation of the flavor group.

The $\Zorb$ projection correlates the states
in the ${\bf (4,4)}$ representation
with operators in even-dimensional
representations of $\SU(2)_L$ \cite{Ennes:2000}.
These open string vacuum states therefore correspond to
the gauge theory operators
with an odd numbers of $A$'s and $B$'s. The natural candidate 
is
\be
\label{SpSpdiff}
  Q_1^{I} J_1 (A_1 B_1)^{n} A_1 Q_2^K\,,
\ee
which has $H_\two=1$ and belongs to the $({\bf 4},{\bf 4})$ representation
of the flavor group.
(The operator $Q_2^K J_2 B_1 (A_1 B_1)^n Q_1^I$
is equivalent to (\ref{SpSpdiff}) using the orientifold projections 
(\ref{ABconstraint}).)

Zero-momentum oscillators in the $w_2$, $\bw_2$ directions of $\tIR^4$
acting on the open string vacuum
correspond to the insertion of the $H_\two=1$ fields
$A_2 B_1 + A_1 B_2 $ and $B^\dagger_1 B_1 + A_1 A^\dagger_1$
into the operators \eqs{SpSpsame} and \eqs{SpSpdiff}.

Zero-momentum oscillators in the $w_3$, $\bw_3$  directions
of $\tIR^4$ (which are the directions transverse to the D7-branes)
are absent from the open string spectrum \cite{Berenstein:2002b},
so the corresponding operators on the gauge theory side 
should be absent as well.
This follows from the fact that, as in section \ref{ssua}, the 
insertion of $\phi_{1,2}$ (which has $H_\two=1$) can be shown 
to give rise to an operator that vanishes by the 
F-term equations.  It is therefore not a chiral
primary operator and is consequently ruled out, in agreement with the
string theory result \cite{Berenstein:2002b}.

\setcounter{equation}{0}
\section{The $\cN=2$ $\Sp{\times}\SU{\times} \cdots{\times} \SU{\times}\Sp$
orientifold}
\label{sgoose2}

The $\sp{2N} {\times}\sp{2N} $ theory discussed in section~\ref{ssp1} is
the first in an infinite  series of conformal models with gauge groups
$\Sp{\times}\SU{\times}\cdots{\times}\SU{\times}\Sp$ \cite{Park:1998}.
The orientifold group for these theories
has the form
$G = \Z^{\mathrm{orb}}_{2k}{\times} \Zori $.
As before, $\Zori = \{ \1,\Om (-1)^{F_L} R_{45} \}$
and $\Z^{\mathrm{orb}}_{2N_2} = \{1,\tha,\ldots,\tha^{2 N_2-1}\}$.
The action of the orbifold on the coordinates transverse to the D3-branes is the same as in section \ref{sgoose1}. For a particular 
choice of the orientifold projection acting on the D3-branes
(different from the one in section~\ref{sgoose1})
the gauge group becomes
\cite{Park:1998}
\be \label{spsusp}
\Sp(v_0)\times \SU(v_1) \times \cdots \times \SU(v_{N_2-1}) \times \Sp(v_{N_2})\,, \ee
and the matter content is (in $\cN=2$ language)
\be
\bigoplus_{j=1}^{N_2} (\fund_{j-1},\bfund_{j})  +
\half w_0\fund_0 + {\displaystyle \bigoplus_{j=1}^{N_2-1}} w_j \fund_{j}
+ \half w_{N_2} \fund_{N_2} \,.
\ee
The $w_j$'s are non-negative integers
constrained by the equation
$\half (w_0 + w_{N_2}) + \sum_{j=1}^{N_2-1} w_j = 4$.
The $v_j$'s obey constraints arising from the
vanishing of the beta-function(s) of the field theory~\cite{Park:1998}.
To leading order in $N_2/ N_1$,
where $N_1$ ($\equiv v_0/2$) equals the rank of the first factor of the product gauge group,
the $v_j$'s corresponding to each of the group factors are
equal and take the value $2N_1$ \cite{Naculich:2001b}, 
and the gauge group is essentially
$\sp{2N_1}\times\su{2N_1}^{N_2-1}\times\sp{2N_1}$. 
We will not make any assumptions about the magnitude of the ratio 
$N_2/N_1$;
see, however, the discussion in section \ref{sgoose1}. 
The quantum numbers of the scalar fields of 
the $\Sp{\times}\SU{\times} \cdots{\times} \SU{\times}\Sp$ 
gauge theory (\ref{spsusp}) are given in the table below.

\vbox{
\begin{center}
\begin{tabular}{|c|c|c|c|c|c|c|c|}
\hline
CFT field & Representation  &
$\JR$  &
$J_R^3$&
$J_L^3$&
$\Delta$  &
$H_\two$  \\
\hline
& & & & & & \\[-13pt]
\hline
& & & & & & \\[-13pt]
$\phi_r$ 	& $\mathrm{adjoint}_r$ 	& $1$ &  0 & 0& 1 & $1$ \\[1pt]
$\phi_r^\dag$ 	& $\mathrm{adjoint}_r$ 	& $-1$ & 0 & 0& 1 & $1$ \\[1pt]
$(A_p,(B^p)^\dag)$ &$(\fund_{p-1},\bfund_p)$&0 & $(\half,-\half)$& $\half$ & 1 & (0,1) \\[1pt]
$(B^p,(A_p)^\dag)$ &$(\bfund_{p-1},\fund_p)$&0 & $(\half,-\half)$&$-\half$ & 1 & (1,2) \\[1pt]
$(Q_r, \tQ_r^\dag) $&$ \half w_r \fund_r$ & 0 &$(\half,-\half) $ & 0 & 1 &
$(\half,\thrhalf)$\\[1pt]
$(\tQ_r, Q_r^\dag)$& $\half w_r \bfund_r$ & 0&$(\half,-\half)$& 0 & 1 &
$(\half,\thrhalf)$\\[3pt]
\hline
\end{tabular}
\end{center}
\begin{center}
{\noindent\footnotesize{\bf Table \arabic{table}:\stepcounter{table}}
$\su{2}_R$ multiplets of scalar fields
in the
$\cN=2$ $\Sp{\times}\SU{\times} \cdots{\times} \SU{\times}\Sp$ theory.}
\end{center}
}

In the table, $r$ takes the values $r=0,\ldots, N_2$,
whereas $p$ ranges from 1 to $N_2$.
The quantum number $J^3_L$ in the table above
denotes a $\U(1)_L$ quantum number; generically there
is no $\SU(2)_L$ symmetry (except when $N_2 =1$).

Scaling $N_1$ and $N_2$ to infinity together in Penrose limit \two{},
as in sec.~\ref{sgoose1},
we obtain a background with a compact null direction,
in which the string theory 
can be quantized using the DLCQ scheme,
as in ref.~\cite{Mukhi:2002}.

\medskip
\noindent {\it Closed string sector}
\medskip

\noindent Since there is no bifundamental representation
connecting the last and first factors of the gauge group,
one cannot construct a gauge invariant operator as a ``loop''
of $H_\two=0$ bifundamental fields.
However, we may define
\be
\cA_a{}^\bb = (A_1 A_2 \cdots  A_{N_2})_{a}{}^{\bb}, \qquad\qquad
a= 1, \cdots, v_0, \qquad \bb = 1, \cdots, v_{N_2}\,,
\ee
and then use the symplectic units $J_1$ and $J_2$, of $\sp{v_0}$ 
and $\sp{v_{N_2}}$ respectively, 
to define
${\mathcal{B}_\ba}{}^b$ by $\mathcal{B} = J_2 \cA^T J_1$.
In terms of these two operators, a  gauge 
invariant $H_\two=0$ operator may
be constructed
\be
\la{moosetwovac}
\tr[(\cA\mathcal{B})^k],
\ee
which may be identified with the closed string ground state.
Here $k$ is an arbitrary non-negative integer which labels 
the DLCQ momentum states as in \cite{Mukhi:2002}.

The pictorial interpretation of this operator is as a ``loop''
around a quiver/moose diagram, with two half-circles 
related via the orientifold projection 
(again reminiscent of the type IIA picture of the gauge theory). 

As in section \ref{sgoose1}, 
one can write the gauge invariant operators corresponding to
the vacuum states \eqs{moosetwovac} as well as the excited states in terms 
of cover space fields. 

\medskip
\noindent {\it Open string sector}
\medskip

\noindent The open string vacuum is
constructed out of a string of $H_\two=0$ operators with two $Q$'s
at the ends, exactly as in section \ref{sgoose1}.
Assuming that $Q^I$ (or $\tQ^I$) for a particular $I$ belongs to the
$r$th factor of the gauge group, we can define
\bea
\cQ^{I}_{a} &=&  (A_{1} \cdots A_{r} Q^I )_{a}\,, \ret\
\tcQ_{I}^{\ba} &=& (\tQ^I A_{r+1} \cdots A_{N_2})^{\ba}\,.
\ee
{}From these, one can construct operators corresponding
to the open string vacua, in analogy with (\ref{SpSpsame}), 
(\ref{SpSpdiff})
\bea
(\cQ^I)^T J_1 \left( \cA \mathcal{B} \right)^{k} \cQ^K\,,\qquad\qquad
\tcQ_I \left( \mathcal{B} \cA \right)^{k} J_2 (\tcQ_K)^T \,,\qquad\qquad
\tcQ_I \mathcal{B} \left( \cA \mathcal{B} \right)^{k} \cQ^K\,.
\eea
Further details can be found in section \ref{sgoose1}.

In addition to the series of models studied in this section and in 
section \ref{sgoose1}, 
there is another infinite series of models containing D7 branes,
with gauge group
$\Sp{\times}\SU {\times}\cdots{\times}\SU{\times}\SU$~\cite{Park:1998}.  
These models can also be treated using the methods of this paper,
and possess both a closed and open string sector in the Penrose limit. 

In ref.~\cite{Park:1998} four additional infinite series of models 
were constructed that do not contain D7-branes. 
It should also be possible to analyze the pp-wave limits of these models.

\setcounter{equation}{0}
\section{$\cN=1$ $\sp{2N}{\times}\sp{2N}$ with $2(\protect\fund,\protect\fund)\oplus 4(1,\protect\fund) \oplus 4(\protect\fund,1)$ } \label{ssp2}

In the previous sections, we have discussed various
$\cN=2$ theories and their associated Penrose limits.
It is of interest to extend these results to
$\cN=1$ theories.
The pp-wave limit associated with the 
duality between the $\cN=1$ $\SU(N){\times}\SU(N)$ theory
and string theory on $\mathit{adS}_5{\times}T^{11}$ \cite{Klebanov:1998} 
was treated in refs.~\cite{Itzhaki:2002,Gomis:2002,Zayas:2002}.
Orbifolds of this
theory and their Penrose limits were discussed in ref.~\cite{Oh:2002}. 
Various $\cN=1$ orbifolds of
$\mathit{adS}_5{\times}S^5$ have been discussed in 
ref.~\cite{Gomis:2002,Floratos:2002,Oh:2002};
recently  non-supersymmetric models have also been 
discussed~\cite{Bigazzi:2002}.

In this section we discuss a particular orientifolded version of
the $\mathit{adS}_5{\times}T^{11}$ model and its Penrose limit.
By studying this model we are able to test the ideas
of ref.~\cite{Berenstein:2002b} in a more complicated example.
The orientifold we study was previously discussed
in ref.~\cite{Naculich:2001b,Naculich:2002a,Schnitzer:2002}.
The field theory dual is an
$\cN=1$ $\Sp(2N){\times}\Sp(2N)$ field theory
with matter content (in $\cN=1$ language)
$2(\fund,\fund)\oplus 4(\fund,1)\oplus 4(1,\fund)$.
As in ref.~\cite{Naculich:2002a} and in sec.~\ref{ssp1},
for notational clarity the bifundamental chiral multiplets
are denoted using the doubled set of fields
${A_{ia}}^\bb$ $(i=1,2)$ and $B^{i~b}_\ba$ $(i=1,2)$,
with the constraint\footnote{In ref.~\cite{Naculich:2002a},
$(B^1, B^2)$ was written as $(B_1, B_2)$.}
\be
\la{ABconstrainttwo}
B^1 = -J_2 A_2^T J_1\,, \qquad B^2 =  J_2 A_1^T J_1\,.
\ee
The chiral multiplets in the $(\fund,1)$ and $(1,\fund)$ representations
will be denoted by $Q^I_{1a}$ and $Q^I_{2\ba}$ ($I=1,\ldots,4)$,
respectively.
For convenience, we also introduce fields
${\tQ_{1I }^{ a}} $ and ${\tQ_{2I }^{\ba}} $, 
related to $Q^I_{1a}$ and $Q^I_{2\ba}$ by \cite{Naculich:2002a}
\be
\tQ_{1I}^T =  - g_{IJ } J_1  Q_1^{J} , \qquad
\tQ_{2I}^T=   - g_{IJ}  J_2  Q_2^{J} ,
\ee
where $g_{IJ}$ is the metric for either of the two factors of the
$\SO(4){\times}\SO(4)$ flavor symmetry group.
The quantum numbers of the scalar fields and their
complex conjugates are displayed in the table 
below.

\begin{center}
\begin{tabular}{|c|c|c|c|c|c||c|c|c|c|c|c|}
\hline Field &
Representation &
$\JR$  &
$J_R^3$&
$\Delta$  &
$H$ &
 Field &
Representation &
$\JR$  &
$J_R^3$&
$\Delta$  &
$H$ \\
\hline
& & & & & & & & & & & \\[-13pt]
\hline
& & & & & & & & & & & \\[-13pt]
$A_1$ 		& $(\fund,\bfund)$ & $ \half$ & $ \half$ & $\tf$ & $0$ &
$(A_1)^\dag$ 	& $(\bfund,\fund)$ & $-\half$ & $-\half$ & $\tf$ & $\thrhalf$ 
\\[1pt]
$A_2$ 		& $(\fund,\bfund)$ & $\half$  & $-\half$ & $\tf$ & $1$ &
$(A_2)^\dag$    & $(\bfund,\fund)$ & $-\half$ & $\half$  & $\tf$ & $\half$ 
\\[1pt]
$B^1$ 		& $(\bfund,\fund)$ & $\half $ & $-\half$ & $\tf$ & $1$ &
$(B^1)^\dag$	& $(\fund,\bfund)$ & $-\half$ & $\half $ & $\tf$ & $\half$ 
\\[1pt]
$B^2$ 		& $(\bfund,\fund)$ & $ \half$ & $\half$  & $\tf$ & $0$ &
$(B^2)^\dag$ 	& $(\fund,\bfund)$ & $-\half$ & $-\half$ & $\tf$ & $\thrhalf$
\\[1pt]
$Q^I_1$  		&$ (\fund, 1)$     & $\half$  & 0 & $\tf$ & $\half$  &
$(Q^I_1)^\dag$  	&$ (\bfund, 1)$    & $-\half$ & 0 & $\tf$ & $\thrhalf$ 
\\[1pt]
$\tQ_{1I}$  	&$ (\bfund, 1)$    & $\half$  & 0 & $\tf$ & $\half$ &
$(\tQ_{1I})^\dag$&$ (\fund, 1)$     & $-\half$  & 0 & $\tf$ & $\thrhalf$ 
\\[1pt]
$Q^I_2$ 		& $(1, \fund)$ 	   & $\half$  & 0 & $\tf$ & $\half$ &
$(Q^I_2)^\dag$  	&$ (1, \bfund)$    & $-\half$ & 0 & $\tf$ & $\thrhalf$ 
\\[1pt]
$\tQ_{2I}$  	&$ (1,\bfund)$ 	   & $\half$  & 0 & $\tf$ & $\half$ &
$(\tQ_{2I})^\dag$&$ (1,\fund)$      & $-\half$  & 0 & $\tf$ & $\thrhalf$ \\[3pt]
\hline
\end{tabular}\label{tablespxsp2}
\end{center}
\begin{center}
{\noindent\footnotesize {\bf Table \arabic{table}:\stepcounter{table}}
Scalar fields
in the $\cN=1$ $\Sp(2N){\times}\Sp(2N)$ theory with
$2(\fund,\fund)\oplus 4(1,\fund)\oplus 4(\fund,1)$.}
\end{center}

The superpotential for the $\cN=1$ $\Sp(2N){\times}\Sp(2N)$ theory is
given by \cite{Naculich:2002a}
\bea
\label{N=1super}
\mathcal{W}_{\cN=1}&=&- \bigg[
\tr (A_1 B^1 A_2 B^2 - B^1 A_1 B^2 A_2)
+ \half \tQ_{1 I} Q_1^{J } \tQ_{1 J } Q_1^{I}
  - \half \tQ_{2 I} Q_2^{J} \tQ_{2 J} Q_2^{I}
\non\\
&&\qquad \quad +\,\tQ_{1 I} (A_1 B^1 + A_2 B^2) Q_1^{I}
- \tQ_{2 I} (B^1 A_1 + B^2 A_2) Q_2^{I}
\bigg] \,.
\eea
{}from which follow the independent F-term equations \cite{Naculich:2002a}
 \bea
\label{ABFterm}
B^1 A_2 B^2 - B^2 A_2 B^1 + B^1 Q_1^I \tQ_{1I}
- Q_2^I\tQ_{2I} B^1 &=&0\,,\non\\
B^2 A_1 B^1 - B^2 A_1 B^2 + B^2 Q_1^I \tQ_{1I}
-  Q_2^I \tQ_{2I} B^2&=&0 \,, \non \\
( A_1 B^1  + A_2 B^2 + Q_1^{K} \tQ_{1K})Q_1^{I} &=& 0 \,, \non \\
 (B^1 A_1 + B^{2}A_2 + Q_2^K \tQ_{2K}) Q_2^{I} &=& 0 \,.
\eea
Additional relations can be obtained by taking the
transposes of eqs.~\eqs{ABFterm}.

The $\cN=1$ $\sp{2N}{\times}\sp{2N}$ gauge theory just discussed
arises as the theory on a stack of D3-branes at a conifold singularity
modded out by a $\Zori$ orientifold projection.
In the near-horizon limit, the geometry becomes
$\mathit{adS}_5 {\times} T^{11}/\Zori$.
The metric on $\mathit{adS}_5$ is given by eq.~(\ref{adS5});
the metric on $T^{11}$ is  \cite{Candelas:1990}
\be
\label{T11met}
\D s_{T^{11}}^2\!
= \frac{1}{9}(\D \psi + \cos\tha_1\D \vphi_1 + \cos \tha_2 \D \vphi_2)^2
\!+ \!\frac{1}{6}(\D \tha_1^2 + \sin^2\tha_1 \D\vphi_1^2)
 + \!\frac{1}{6}(\D \tha_2^2 + \sin^2\tha_2 \D\vphi_2^2).
\ee
The orientifold acts by interchanging the two
spheres \cite{Naculich:2001b,Naculich:2002a}:
$(\vphi_1,\tha_1)\leftrightarrow (\vphi_2,\tha_2)$.
The fixed point set $\tha_1=\tha_2$, $\vphi_1=\vphi_2$
represents the position of the O7-plane and D7-branes.

\bigskip
\noindent{\bf The Penrose limit}
\medskip

\noindent Parameterizing the angular variables in eq.~\eqs{T11met} as
\cite{Itzhaki:2002,Gomis:2002,Zayas:2002}
\be
\vphi_i = \xi - \chi_i\,, \qquad
\psi = \xi + \chi_1 + \chi_2\,, \qquad
\tha_i = \frac{\sqrt{6}}{R} y_i\,,
\ee
introducing
\be
x^+ = \half (t + \xi)\,, \qquad
x^- = \half R^2 (t - \xi)\,, \qquad
r   = R \rho\,,
\ee
and then taking the $R\rar \infty$ limit gives the standard pp-wave metric:
\bea
\D s^2 &=&  R^2 \left( \D s_{\mathit{adS}_5}^2 + \D s_{T^{11}}^2  \right)\,,\\
& \to &-4 \D x^+ \D x^- - (r^2 + y_1^2 + y_2^2) (\D x^+)^2
 + \D r^2 + r^2 \D \Om_3^2
 + \D y_1^2 + y_1^2 \D \chi_1^2
 + \D y_2^2 + y_2^2 \D \chi_2^2.\non
\ee
The orientifold survives the limit
and acts by interchanging $(y_1,\chi_1)$ and $(y_2,\chi_2)$.
The D7-branes also survive the limit and are located at
the fixed point of the orientifold, i.e.~at $y_1=y_2$, $\chi_1=\chi_2$.
This gives rise to an open-string sector of the
string theory in the Penrose limit.
(Contrary to the situation for the $\cN=2$ models,
there does not appear to exist a limit in which the
D7-branes disappear.)

The limiting metric and the orientifold action are identical
(after a trivial coordinate redefinition) to the ones
considered in ref.~\cite{Berenstein:2002b} (the gauge group on
the D7-branes is  different though).
Thus the string theory analysis carried out in ref.~\cite{Berenstein:2002b}
carries over essentially unchanged to our case,
but the field theory analysis is different.

The $\cN=1$ $\sp{2N}{\times}\sp{2N}$ gauge theory
has a $\U(1)_R{\times}\SU(2){\times}\SU(2)$ symmetry;
on the string theory side, this  arises from the isometry group of $T^{11}$.
We use the following notation:
$J_R$ is the generator of $\U(1)_R$,
whereas $J^3_{R}$ is sum of the Cartan generators of the two $\SU(2)$'s.
Hence,
\be
J^3_R = -i\partial_{\vphi_1} -i\partial_{\vphi_2},\qquad
\JR = -2i\partial_{\psi}\,.
\ee
For the Penrose limit considered above,
we have $H = \De - J$ where
$J = \half J_R + J^3_R$ \cite{Itzhaki:2002,Gomis:2002,Zayas:2002}.

\medskip
\noindent{\it
Closed string sector}
\medskip

\noindent The ground state in the closed string sector
corresponds (in our conventions) to the $H=0$ gauge theory
operator
\be
\la{N=1vacuum}
\tr(B^2A_1 )^n\,.
\ee
Operators with $H=1$
are $D_{\mu}(B^2A_1 )$,
$B^1A_1 $, $B^2A_2 $,
$(A_2)^\dag A_1 $ and $B^2(B^1)^\dag $.
(It was pointed out in ref.~\cite{Itzhaki:2002}
that the conformal dimension of the latter two
operators is not the naive sum of the dimensions of the constituents,
but is $\De = 2$, resulting in $H=1$.)
In the unorientifolded case,
insertions of these operators in \eqs{N=1vacuum}
correspond to oscillators acting on the
vacuum of the string theory \cite{Itzhaki:2002,Gomis:2002,Zayas:2002}.

In the orientifolded theory, the zero-momentum oscillator in the direction
transverse to the orientifold plane is odd under $\Zori$,
and so single oscillator states in this direction are projected out.
Correspondingly, a single insertion of the operator corresponding
to this direction yields a vanishing operator,
as we will now show.
The operators $N_{ij} = B_{\ba}^{j~c} A_{i\,c}{}^\bb$
satisfy the orientifold projections \cite{Naculich:2002a}
\bea
\label{Nconstraint}
N_{11} =  J_2 N_{22}^T J_2\,, \qquad \qquad
N_{12} &=& -J_2 N_{12}^T J_2 \,, \qquad \qquad
N_{21} = -J_2 N_{21}^T J_2 \,,
\eea
by virtue of eqs.~\eqs{ABconstrainttwo}.
It follows that $(N_{11}+N_{22})J_2$ is symmetric whereas
$N_{12}J_2$, $N_{21}J_2$ and $(N_{11}-N_{22})J_2$ are antisymmetric.
This implies that the number of insertions of
$N_{11}+N_{22}$ into the vacuum state $\tr(N_{12})^n$ must be even.
This result corresponds to the fact that the coordinate on the
string theory side corresponding to $(N_{11}+N_{22})$
changes sign under the orientifold \cite{Naculich:2002a}.
Similarly, the other operator which changes sign under the orientifold
is $(A_2)^\dag A_1 + B^2(B^1)^\dag $;
only an even number of insertions of this operator is possible.
This follows from the fact that
$[(A_2)^\dag A_1 + B^2(B^1)^\dag] J_2$ is symmetric
due to eq.~(\ref{ABconstrainttwo}).
We have thus shown that the field theory reproduces the string theory
result.

\medskip
\noindent{\it Open string sector}
\medskip

\noindent
There are three possible open string vacua
formed from an $H=0$ operator with two ``quarks'' at the ends:
\be \label{osvs}
Q^{[I}_{1} J_1 (A_1 B^2 )^n Q^{K]}_{1}  \,, \qquad \quad
Q^{[I}_{2} J_2 (B^2 A_1)^n Q^{K]}_{2}\,, \qquad  \quad
Q^I_{1} J_1 (A_1 B^2)^n A_1 Q^K_{2}\,.
\ee
Because of the antisymmetry of
$J_2 (B^2 A_1)^n $  and $J_1 (A_1 B^2)^n $,
the first two operators in \eqs{osvs} are antisymmetric
in the interchange of $I$ and $K$,
and therefore belong to the adjoint ${\bf 6}$ of $\SO(4)$.
This is consistent with the fact that the flavor symmetry group
for the D7's is $\SO(4){\times}\SO(4)$,
with the first two operators in \eqs{osvs}
corresponding to the two independent $\SO(4)$ factors
(two stacks of D7-branes).
The last operator in \eqs{osvs} belongs to
the $({\bf 4},{\bf 4})$ representation of $\SO(4){\times}\SO(4)$
and represents the vacuum states for strings
connecting the two stacks of D7-branes.

As in the closed string sector, excitations of the
open string correspond to insertions of $H=1$ operators
into the open string  vacua \eqs{osvs}.
Operators corresponding to insertions of zero-momentum
oscillators  in the directions transverse to the D7-branes 
should be absent however \cite{Berenstein:2002b}.
As an example of this, consider the insertion of
$N_{11} + N_{22}$ into $Q^I_{2} J_2 N_{12}^n  Q^K_{2}$.
We will now show that such 
an operator vanishes by using the F-term equations.
{}From the F-term equations (\ref{ABFterm}),
it follows that $N_{11} + N_{22} + Q_2^I \tQ_{2I}$
commutes with all the $N_{ij}$'s.
Therefore $N_{11}+N_{22}$ can be commuted past
$N_{12}$ modulo a $Q^2$ term.
This $Q^2$ term splits the
operator into a ``multiple trace'' operator of the form
$[Q_2 \cdots Q_2][Q_2 \cdots Q_2]$,
which is subleading in the ${1}/{N}$ expansion.
(It is natural to interpret this as a splitting
of the open string into two.)
Thus, $N_{11}+N_{22}$ can be moved to the end of the operator,
where, again by the F-term equations \eqs{ABFterm},
it gives zero when acting on the $Q_2$
(again modulo a $Q^2$ term, which gives a subleading
contribution).
Since the operator vanishes (modulo subleading terms)
when using the F-term equations it is not a protected operator,
consequently the
corresponding zero-momentum oscillator is absent.
Similarly, we expect that an insertion
of $(A_2)^\dag A_1 + B^2(B^1)^\dag $ should give rise
to an unprotected operator, 
reflecting the absence of the corresponding string zero-momentum state.
Thus the field theory reproduces the expected string theory
result \cite{Berenstein:2002b}.

\section{Summary} \label{ssum}
In this paper we have studied the pp-wave limits of a variety of elliptic
models with O7-planes and D7-branes, as well as the pp-wave limit of a
particular orientifold of $\mathit{adS}_5{\times}T^{11}$. The
theories with D7-branes have a Penrose limit with both open and closed string
sectors.

For most cases, two
different Penrose limits were discussed: limit A, which gives
an orbifolded pp-wave and which thus contains both an untwisted and
a twisted sector; and limit B, which gives an orientifold of the
pp-wave with D7-branes, and which thus contains both closed and
open string sectors. In limit A (B) a remnant of the
orientifold (orbifold) part of the full orientifold group is
manifested in the classification of the chiral and ``near-chiral''
primaries of the theories.

The two infinite series of elliptic models discussed in sections
\ref{sgoose1} and \ref{sgoose2}
allow pp-wave scaling limits analogous to the one discussed
in refs.~\cite{Mukhi:2002,Alishahiha:2002b},
in which there is a compact light-cone null direction.
The models that we discuss contain both open and closed strings.

In section \ref{ssp2} by studying a particular orientifold
of $\mathit{adS}_5{\times}T^{11}$ we were able to test the ideas
of ref.~\cite{Berenstein:2002b} in a more complicated example and agreement
was found between the string and field theory results.

We have not touched upon the important topic of interactions, 
but it would clearly be of interest to gain a better understanding of 
the interactions between closed and open strings and to elucidate 
the splitting of open strings. 
For some recent work on interactions in 
the pp-wave background and in the dual gauge theory,
see e.g. refs.~\cite{Kristjansen:2002,Constable:2002,Gopakumar:2002,Int}.

\section*{Acknowledgement}
HJS would like to thank the string theory group and Physics Department of Harvard University for their hospitality extended over a long period of time.

\begingroup\raggedright\endgroup


\begin{thebibliography}{10}

\bibitem{Maldacena:1998}
J.~M. Maldacena, ``The large $N$ limit of superconformal field theories and
  supergravity.'' Adv. Theor. Math. Phys. {\bf 2} (1998) 231--252,
  {{\tt
  hep-th/9711200}}; 
S.~S. Gubser, I.~R. Klebanov, and A.~M. Polyakov, ``Gauge theory correlators
  from non-critical string theory.'' Phys. Lett. {\bf B428} (1998) 105--114,
  {{\tt
  hep-th/9802109}}; 
E.~Witten, ``Anti-de Sitter space and holography.'' Adv. Theor. Math. Phys.
  {\bf 2} (1998) 253--291,
  {{\tt
  hep-th/9802150}}. 

\bibitem{Berenstein:2002a}
D.~Berenstein, J.~M. Maldacena, and H.~Nastase, ``Strings in flat space and pp
  waves from $\cN = 4$ super Yang Mills.'' JHEP {\bf 04} (2002) 013,
  {{\tt
  hep-th/0202021}}. 

\bibitem{Gueven:2000}
R.~G{\"u}ven, ``Plane wave limits and T-duality.'' Phys. Lett. {\bf B482}
  (2000) 255--263,
  {{\tt
  hep-th/0005061}}. 

\bibitem{Blau:2001}
M.~Blau, J.~Figueroa-O'Farrill, C.~Hull, and G.~Papadopoulos, ``A new maximally
  supersymmetric background of IIB superstring theory.'' JHEP {\bf 01} (2002)
  047, {{\tt
  hep-th/0110242}}; 
M.~Blau, J.~Figueroa-O'Farrill, C.~Hull, and G.~Papadopoulos, ``Penrose limits
  and maximal supersymmetry.'' Class. Quant. Grav. {\bf 19} (2002) L87--L95,
  {{\tt
  hep-th/0201081}}. 

\bibitem{Metsaev:2001}
R.~R. Metsaev, ``Type IIB Green-Schwarz superstring in plane wave 
Ramond-Ramond background.'' Nucl. Phys. {\bf B625} (2002) 70--96,
  {{\tt
  hep-th/0112044}}. 

\bibitem{Penrose:1976}
R.~Penrose, ``Any spacetime has a plane wave as a limit.'' in {\em Differential
  Geometry and Relativity}.
\newblock Reidel, Dordrecht, 1976.

\bibitem{Metsaev:2002}
R.~R. Metsaev and A.~A. Tseytlin, ``Exactly solvable model of superstring in
  plane wave Ramond-Ramond background.''
  {{\tt
  hep-th/0202109}}. 

\bibitem{Berenstein:2002b}
D.~Berenstein, E.~Gava, J.~M. Maldacena, K.~S. Narain, and H.~Nastase, ``Open
  strings on plane waves and their Yang-Mills duals.''
  {{\tt
  hep-th/0203249}}. 

\bibitem{Open}
M.~Bill{\'o} and I.~Pesando, ``Boundary states for GS superstrings in an Hpp wave
  background.''
  {{\tt
  hep-th/0203028}}; 
C.-S. Chu and P.-M. Ho, ``Noncommutative D-brane and open string in pp-wave
  background with B-field.''
  {{\tt
  hep-th/0203186}}; 
A.~Dabholkar and S.~Parvizi, ``Dp branes in pp-wave background.''
  {{\tt
  hep-th/0203231}}; 
P.~Lee and J.-w. Park, ``Open strings in PP-wave background from defect
  conformal field theory.''
  {{\tt
  hep-th/0203257}}; 
A.~Kumar, R.~R. Nayak, and Sanjay, ``D-brane solutions in pp-wave background.''
  {{\tt
  hep-th/0204025}}; 
D.-s. Bak, ``Supersymmetric branes in PP wave background.''
  {{\tt
  hep-th/0204033}}; 
K.~Skenderis and M.~Taylor, ``Branes in AdS and pp-wave spacetimes.''
  {{\tt
  hep-th/0204054}}; 
V.~Balasubramanian, M.-x. Huang, T.~S. Levi, and A.~Naqvi, ``Open strings from
  $\cN = 4$ super Yang-Mills.''
  {{\tt
  hep-th/0204196}}; 
Y.~Imamura, ``Large angular momentum closed strings colliding with D-branes.''
  {{\tt
  hep-th/0204200}}; 
H.~Takayanagi and T.~Takayanagi, ``Open strings in exactly solvable model of
  curved space-time and PP-wave limit.'' 
  {{\tt
  hep-th/0204234}}; 
P.~Bain, P.~Meessen, and M.~Zamaklar, ``Supergravity solutions for D-branes in
  Hpp-wave backgrounds.''
  {{\tt
  hep-th/0205106}}; 
M.~Alishahiha and A.~Kumar, ``D-brane solutions from new isometries of
  pp-waves.''
  {{\tt
  hep-th/0205134}}; 
O.~Bergman, M.~R. Gaberdiel, and M.~B. Green, ``D-brane interactions in type
  IIB plane-wave background.''
  {{\tt
  hep-th/0205183}}; 
S.~Seki, ``D5-brane in Anti-de Sitter space and Penrose limit.''
  {{\tt
  hep-th/0205266}}; 
D.~Mateos and S.~Ng, ``Penrose limits of the baryonic D5-brane.''
  {{\tt
  hep-th/0205291}}; 
S.~S. Pal, ``Solution to worldvolume action of D3 brane in pp-wave
  background.''
  {{\tt
  hep-th/0205303}}. 

\bibitem{Gukov:1998}
S.~Gukov and A.~Kapustin, ``New $\cN = 2$ superconformal field theories from
  M/F theory orbifolds.'' Nucl. Phys. {\bf B545} (1999) 283--308,
  {{\tt
  hep-th/9808175}}. 

\bibitem{Ennes:2000}
I.~P. Ennes, C.~Lozano, S.~G. Naculich, and H.~J. Schnitzer, ``Elliptic models,
  type IIB orientifolds and the AdS/CFT correspondence.'' Nucl. Phys. {\bf
  B591} (2000) 195--226,
  {{\tt
  hep-th/0006140}}. 

\bibitem{Park:1998}
J.~Park and A.~M. Uranga, ``A note on superconformal $\cN = 2$ theories and
  orientifolds.'' Nucl. Phys. {\bf B542} (1999) 139--156,
  {{\tt
  hep-th/9808161}}. 

\bibitem{Mukhi:2002}
S.~Mukhi, M.~Rangamani, and E.~Verlinde, ``Strings from quivers, membranes from
  moose.'' JHEP {\bf 05} (2002) 023,
  {{\tt
  hep-th/0204147}}. 

\bibitem{Alishahiha:2002b}
M.~Alishahiha and M.~M.~Sheikh-Jabbari,
``Strings in PP-waves and worldsheet deconstruction,''
  {{\tt hep-th/0204174}}.  

\bibitem{Blau:2002b}
M.~Blau, J.~Figueroa-O'Farrill, and G.~Papadopoulos, ``Penrose limits,
  supergravity and brane dynamics.''
  {{\tt
  hep-th/0202111}}. 

\bibitem{Witten:1998b}
E.~Witten, ``Baryons and branes in anti de Sitter space.'' JHEP {\bf 07} (1998)
  006, {{\tt hep-th/9805112}}.

\bibitem{Floratos:2002}
E.~Floratos and A.~Kehagias, ``Penrose limits of orbifolds and orientifolds.''
  {{\tt
  hep-th/0203134}}. 

\bibitem{Sen:1996}
A.~Sen, ``F-theory and Orientifolds.'' Nucl. Phys. {\bf B475} (1996) 562--578,
  {{\tt
  hep-th/9605150}}; 
T.~Banks, M.~R. Douglas, and N.~Seiberg, ``Probing F-theory with branes.''
  Phys. Lett. {\bf B387} (1996) 278--281,
  {{\tt hep-th/9605199}};
O.~Aharony, J.~Sonnenschein, S.~Yankielowicz, and S.~Theisen, ``Field theory
  questions for string theory answers.'' Nucl. Phys. {\bf B493} (1997)
  177--197,
  {{\tt
  hep-th/9611222}}; 
M.~R. Douglas, D.~A. Lowe, and J.~H. Schwarz, ``Probing F-theory with multiple
  branes.'' Phys. Lett. {\bf B394} (1997) 297--301,
  {{\tt hep-th/9612062}}.

\bibitem{Fayyazuddin:1998}
A.~Fayyazuddin and M.~Spalinski, ``Large $N$ superconformal gauge theories and
  supergravity orientifolds.'' Nucl. Phys. {\bf B535} (1998) 219--232,
  {{\tt
  hep-th/9805096}}; 
O.~Aharony, A.~Fayyazuddin, and J.~M. Maldacena, ``The large $N$ limit of $\cN
  = 2,1$ field theories from three- branes in F-theory.'' JHEP {\bf 07} (1998)
  013, {{\tt
  hep-th/9806159}}. 

\bibitem{Kim:2002}
N.-w. Kim, A.~Pankiewicz, S.-J. Rey, and S.~Theisen, ``Superstring on pp-wave
  orbifold from large-$N$ quiver gauge theory.''
  {{\tt
  hep-th/0203080}}. 

\bibitem{Douglas:1996}
M.~R. Douglas and G.~W. Moore, ``D-branes, Quivers, and ALE Instantons.''
  {{\tt
  hep-th/9603167}}. 

\bibitem{Kristjansen:2002}
C.~Kristjansen, J.~Plefka, G.~W. Semenoff, and M.~Staudacher, ``A new
  double-scaling limit of $\cN = 4$ super Yang-Mills theory and PP-wave
  strings.''
  {{\tt
  hep-th/0205033}}. 

\bibitem{Constable:2002}
N.~R. Constable {\em et.~al.}, ``PP-wave string interactions from perturbative
  Yang-Mills theory.''
  {{\tt
  hep-th/0205089}}. 

\bibitem{Itzhaki:2002}
N.~Itzhaki, I.~R. Klebanov, and S.~Mukhi, ``PP wave limit and enhanced
  supersymmetry in gauge theories.'' JHEP {\bf 03} (2002) 048,
  {{\tt
  hep-th/0202153}}. 

\bibitem{Alishahiha:2002a}
M.~Alishahiha and M.~M. Sheikh-Jabbari, ``The PP-wave limits of orbifolded
  $\mathit{AdS}_5{\times}S^5$.''
  {{\tt
  hep-th/0203018}}. 

\bibitem{Takayanagi:2002}
T.~Takayanagi and S.~Terashima, ``Strings on orbifolded pp-waves.''
  {{\tt
  hep-th/0203093}}. 

\bibitem{Naculich:2001b}
S.~G. Naculich, H.~J. Schnitzer, and N.~Wyllard, ``$1/N$ corrections to
  anomalies and the AdS/CFT correspondence for orientifolded $\cN = 2$ orbifold
  models and $\cN = 1$ conifold models.''
  {{\tt
  hep-th/0106020}}. 


\bibitem{Susskind:1997}
L.~Susskind, ``Another conjecture about M(atrix) theory.''
  {{\tt
  hep-th/9704080}}. 

\bibitem{Cvetic:2002}
M.~Cveti{\v c}, H.~L{\"u}, and C.~N. Pope, ``Penrose limits, pp-waves and
  deformed M2-branes.''
  {{\tt
  hep-th/0203082}}; 
J.~Michelson, ``(Twisted) toroidal compactification of pp-waves.''
  {{\tt
  hep-th/0203140}}. 

\bibitem{Gopakumar:2002}
R.~Gopakumar, ``String interactions in PP-waves.''
  {{\tt
  hep-th/0205174}}. 

\bibitem{Naculich:2002a}
S.~G. Naculich, H.~J. Schnitzer, and N.~Wyllard, ``A cascading $\cN = 1$
  $\Sp(2N{+}2M){\times}\Sp(2N)$ gauge theory.''
  {{\tt
  hep-th/0204023}}. 

\bibitem{Klebanov:1998}
I.~R. Klebanov and E.~Witten, ``Superconformal field theory on threebranes at a
  Calabi-Yau singularity.'' Nucl. Phys. {\bf B536} (1998) 199--218,
  {{\tt
  hep-th/9807080}}. 

\bibitem{Gomis:2002}
J.~Gomis and H.~Ooguri, ``Penrose limit of $\cN = 1$ gauge theories.''
  {{\tt
  hep-th/0202157}}. 

\bibitem{Zayas:2002}
L.~A. {Pando Zayas} and J.~Sonnenschein, ``On Penrose limits and gauge
  theories.'' JHEP {\bf 05} (2002) 010,
  {{\tt
  hep-th/0202186}}. 

\bibitem{Oh:2002}
K.~Oh and R.~Tatar, ``Orbifolds, Penrose limits and supersymmetry
  enhancement.''
  {{\tt
  hep-th/0205067}}. 

\bibitem{Bigazzi:2002}
F.~Bigazzi, A.~L. Cotrone, L.~Girardello, and A.~Zaffaroni, ``pp-wave and
  non-supersymmetric gauge theory.''
  {{\tt
  hep-th/0205296}}. 

\bibitem{Schnitzer:2002}
H.~J. Schnitzer and N.~Wyllard, 
``An orientifold of $\mathit{adS}_5 {\times} T^{11}$ with D7-branes, 
the associated $\alpha^{\prime 2}$-corrections,
and their role in the dual $\cN=1$ Sp($2N+2M)\times$ Sp($2N$) gauge theory.'' 
{{\tt hep-th/0206071}}. 

\bibitem{Candelas:1990}
P.~Candelas and X.~C. de~la Ossa, ``Comments on conifolds.'' Nucl. Phys. {\bf
  B342} (1990) 246--268. 


\bibitem{Int}
M.~Spradlin and A.~Volovich, ``Superstring interactions in a pp-wave
  background.''
  {{\tt
  hep-th/0204146}}; 
A.~Parnachev and D.~A. Sahakyan, ``Penrose limit and string quantization in
  $\mathit{AdS}_3{\times}S^3$.''
  {{\tt
  hep-th/0205015}}; 
D.~J. Gross, A.~Mikhailov, and R.~Roiban, ``Operators with large R charge in
  $\cN = 4$ Yang-Mills theory.''
  {{\tt
  hep-th/0205066}}; 
C.-S. Chu, P.-M. Ho, and F.-L. Lin, ``Cubic string field theory in pp-wave
  background and background independent Moyal structure.''
  {{\tt
  hep-th/0205218}}; 
Y.-j. Kiem, Y.-b. Kim, S.-m. Lee, and J.-m. Park, ``pp-wave/Yang-Mills
  correspondence: An explicit check.''
  {{\tt
  hep-th/0205279}}; 
M.-x. Huang, ``Three point functions of $\cN=4$ Super Yang Mills from light
  cone string field theory in pp-wave.''
  {{\tt
  hep-th/0205311}}; 
C.-S. Chu, V.~V. Khoze, and G.~Travaglini, ``Three-point functions in $\cN=4$
  Yang-Mills theory and pp-waves.''
  {{\tt hep-th/0206005}}. 

\end{thebibliography}
\end{document}